\documentclass[english,aps]{revtex4}
\usepackage[T1]{fontenc}
\usepackage[latin9]{inputenc}
\usepackage{verbatim}
\usepackage{amsmath}
\usepackage{amssymb}
\usepackage{graphicx}
\usepackage{esint}

\makeatletter

\providecommand{\tabularnewline}{\\}

\@ifundefined{textcolor}{}
{%
 \definecolor{BLACK}{gray}{0}
 \definecolor{WHITE}{gray}{1}
 \definecolor{RED}{rgb}{1,0,0}
 \definecolor{GREEN}{rgb}{0,1,0}
 \definecolor{BLUE}{rgb}{0,0,1}
 \definecolor{CYAN}{cmyk}{1,0,0,0}
 \definecolor{MAGENTA}{cmyk}{0,1,0,0}
 \definecolor{YELLOW}{cmyk}{0,0,1,0}
 }
\makeatother

\usepackage{babel}
\begin{document}

\title{Confinement in a Higgs Model on $R^{3}\times S^{1}$}

\author{Hiromichi Nishimura}
\email[ ]{hnishimura@physics.wustl.edu}
\author{Michael C. Ogilvie}
\email[] {mco@physics.wustl.edu}

\affiliation{Dept. of Physics, Washington University, St. Louis, MO 63130 USA}
\begin{abstract}
We determine the phase structure of an $SU(2)$ gauge theory with
an adjoint scalar on $R^{3}\times S^{1}$ using semiclassical methods.
There are two global symmetries: a $Z(2)_{H}$ symmetry associated
with the Higgs field and a $Z(2)_{C}$ center symmetry associated
with the Polyakov loop in the compact direction. The order
of the deconfining phase transition 
can be either second-order or first-order for $SU(2)$,
depending on the deformation used.
After finding order parameters for the
global symmetries, we show that there are four distinct phases: a
deconfined phase, a confined phase, a Higgs phase, and a mixed confined
phase. The mixed confined phase occurs where one might expect a phase
in which there is both confinement and the Higgs mechanism, but the
behavior of the order parameters distinguishes the two phases. In
the mixed confined phase, the $Z(2)_{C}\times Z(2)_{H}$ global symmetry
breaks spontaneously to a $Z(2)$ subgroup that acts nontrivially
on both the scalar field and the Polyakov loop. We find
explicitly the BPS and KK monopole solutions of the Euclidean field
equations in the BPS limit; these monopoles are extensions of similar
pure gauge theory solutions, where they are constituents of instantons.
In the mixed phase, a linear combination of the Higgs field $\phi$ and $A_{4}$,
the component of the gauge field in the compact direction, enters
into the monopole solutions. In all four phases, Wilson loops orthogonal
to the compact direction are expected to show area-law behavior. We
show that this confining behavior can be attributed to a dilute monopole
gas in a broad region that includes portions of all four phases. The
dilute monopole gas picture breaks down when
the action of a BPS monopole is zero. A duality argument
similar to that applied recently \cite{Poppitz:2011wy} to the Seiberg-Witten model
on $R^3 \times S^1$
shows that the monopole gas picture, arrived at using Euclidean instanton
methods, can be interpreted as a gas of finite-energy dyons. 
\end{abstract}
\thanks{The authors thank the US Department of Energy for support.}
\maketitle

\section{Introduction}

One of the most fundamental questions we can ask about a gauge theory
is its phase diagram. In the standard model, we have seen three fundamentally
different types of behavior: the familiar Coulomb behavior associated
with the massless photon, the Higgs mechanism, and the confinement of quarks
and gluons. These properties are characteristics of different phases:
QCD is in a confined phase at zero temperature and density, while the electroweak
sector of the standard model combines Coulomb and Higgs phases. 

As shown by 't Hooft \cite{'tHooft:1977hy,'tHooft:1979uj}, there
is a fundamental conflict between the Higgs mechanism and confinement.
There is a simple picture of this conflict based on the dual superconductor
picture of confinement. In a type II superconductor, magnetic monopoles
would be confined by magnetic flux tubes, which we interpret as the
Higgs mechanism leading to the confinement of magnetic charges. If
the confined phase of a gauge theory can be interpreted as a dual condensate
of magnetic monopoles, then confinement of non-Abelian electric charge
would follow. 

We will study below the phase structure of an $SU(2)$ adjoint Higgs
model on $R^{3}\times S^{1}$. Together with the scalar potential, a deformation term added to the model 
will allow us to explore what turns out
to be a very rich phase structure. 
The use of $R^{3}\times S^{1}$ with a small circumference,
as opposed to $R^4$, makes the gauge coupling small.
One-loop perturbation theory shows that the
deformation term can be used to move between confined and
deconfined phases.
This in turn allows the study of the interplay
between confinement and the Higgs mechanism
using semiclassical methods.
This model extends recent work
on gauge theories that are confining on $R^{3}\times S^{1}$ for small
circumference $L$ \cite{Myers:2007vc,Unsal:2007vu}.
Typically, we associate this geometry with finite
temperature, and $L$ with the inverse temperature $\beta$, and we
would expect a high-temperature, deconfined phase for small $\beta$.
Recently, methods have been found to change this result: gauge models
have been found where confinement can be understood analytically
at small $L$ using semiclassical methods. The starting point is typically
a gauge theory on $R^{3}\times S^{1}$; a small circumference $L$
for the compact direction implies a small coupling constant $g(L)$
provided $L\Lambda\ll1$, where $\Lambda$ is the characteristic renormalization
group-invariant mass scale of the theory. Such a gauge theory is generally
found in the deconfined phase for small $L$, so it is necessary to
modify the gauge action in order to obtain a confined phase. 
In previous work on deformed gauge theories without fundamental scalars,
there has been good evidence that the confined and deconfined phases
on $R^{3}\times S^{1}$ for small $L$ are continuously connected to
the same phases at large $L$ \cite{Myers:2007vc}.
%Methods to do this are now known, and it is possible to move between
%the confined and deconfined phases in a controlled manner.

The center symmetry associated with the gauge field, which is a $Z(N)_{C}$
symmetry for $SU(N)$, is crucial to our modern understanding of confinement
and deconfinement. The Polyakov loop operator $P\left(\vec{x}\right)$
will be central to our analysis. It is defined as a Wilson loop traversing
a topologically nontrivial path in the compact direction given
by
\begin{equation}
P\left(\vec{x}\right)=\mathcal{P}\,\exp\left[ig\int_{0}^{L}dx_{4}A_{4}\left(\vec{x},x_{4}\right)\right]
\end{equation}
where $\mathcal{P}$ indicates path ordering and $A_{\mu}$ is the
gauge field. It transforms as $P\left(\vec{x}\right)\rightarrow g\left(\vec{x},0\right)P\left(\vec{x}\right)g^{\dagger}\left(\vec{x},0\right)$
under a gauge transformation $g\left(\vec{x},x_4\right)$ so that $Tr_{R}P^{n}\left(\vec{x}\right)$
is gauge-invariant for any representation $R$ and any integer $n$.
The Polyakov loop transforms nontrivially under center symmetry.
For $SU(N)$, this is a transformation that takes 
$Tr_{R}P^{n}\left(\vec{x}\right)$  into $z^{n}Tr_{R}P^{n}\left(\vec{x}\right)$ 
where $z\in Z(N)_{C}$. In a pure gauge
theory at small $L$, the one-loop effective potential for $P$ is
reliable, and indicates that $Z(N)_{C}$ symmetry is spontaneously
broken: The deconfined phase is preferred in this region.
In order to restore the confined, $Z(N)_{C}$-symmetric phase at small
$L$, additional contributions to the effective potential must be
present. Two methods are known, one using adjoint fermions, and the other
a deformation of the gauge action. 
The addition of adjoint representation
fermions to $SU(N)$ gauge theories preserves the global $Z(N)_{C}$
symmetry of the action. With normal antiperiodic boundary conditions
for the fermions, the perturbative effective action for the Polyakov
loop shows that the deconfined phase remains favored at high temperature,
as in the pure gauge case. With periodic boundary conditions for the
fermions, however, this class of field theories can avoid the transition
to the deconfined phase found in the pure gauge theory for sufficiently
light fermion mass and small $L$
\cite{Kovtun:2007py,Myers:2009df,Meisinger:2009ne}.
If the number of adjoint
Dirac fermion flavors $N_{f}$ is less than $11/2$, these systems are asymptotically
free at small $L$, and therefore the effective potential for
$P$ is calculable using perturbation theory. An alternative approach
which is closely related is to add to the gauge action deformation
terms which are local in the noncompact directions, but nonlocal
in the compact direction \cite{Myers:2007vc, Unsal:2008ch}.
Because these terms must respect center
symmetry, they are often referred to as double-trace deformations,
reflecting the fact that $Tr_{A}P=Tr_{F}P^{\dagger}Tr_{F}P-1$. A
minimal choice for the deformation term $S_{d}$, which is adequate
for $SU(2)$ and $SU(3)$, takes the form
\begin{equation}
S_{d}=L\int d^{3}x\,\frac{h_{1}}{L^{4}}\left|Tr_{F}P\left(\vec{x}\right)\right|^{2}
\end{equation}
which favors the confined phase with $Tr_{F}P=0$ for $h_1>0$. For
$N\ge4$, it is necessary to include additional terms to avoid partially
confined phases, as in the case of $SU(4)$ where $Z(4)$ can break
spontaneously to $Z(2)$. In this more general case, the deformation
may be taken to be 
\begin{equation}
S_{d}=L\int d^{3}x\,\sum_{k=1}^{\left[\frac{N}{2}\right]}\frac{h_{k}}{L^{4}}\left|Tr_{F}P^{k}\left(\vec{x}\right)\right|^{2}
\end{equation}
with the confined phase regained at small $L$ if all the $h_{k}$'s
are sufficiently positive.

The change of the action away from that of a pure gauge theory restores
center symmetry in the compact direction in such a way that perturbation
theory can be used to calculate fundamental quantities associated
with the Polyakov loops such as string tensions. In contrast, the
maintenance of center symmetry in the noncompact directions, which
holds for all values of $L$, is nonperturbative. String tensions
are measured by Wilson loops in planes orthogonal to the compact direction.
The mechanism by which the Wilson loop string tension arises is monopole
condensation, via a mechanism first discussed by Polyakov \cite{Polyakov:1976fu}
in the context of a $d=3$ Higgs model. Once the center symmetry is
restored by either method discussed above, the gauge field in the
compact direction, $A_{4}$ automatically acquires a nonzero vacuum
expectation value in an appropriately chosen gauge. 
$A_{4}$ then behaves like a scalar field in the
usual Higgs mechanism, where $SU(N)$ spontaneously breaks down to
$U(1)^{N-1}$. Unlike the case of conventional scalar fields, there
are $N$ monopoles in this case; $N-1$ BPS monopoles and one additional
monopole, called the Kaluza-Klein monopole due to the fact that the fourth
direction is compactified \cite{Kraan:1998kp,Kraan:1998pm,Lee:1998bb}.
Following the work by Polyakov, Unsal and Yaffe were able
to analytically calculate the string tension for the case of $SU(2)$
using the dilute gas approximation of monopoles \cite{Unsal:2008ch}.
A similar result holds for $SU(3)$, although
unfortunately the more general case of $SU(N)$ is not as
tractable \cite{Meisinger:2009ne}.

With confinement in the pure gauge theory on $R^{3}\times S^{1}$
under analytic control, we can now introduce an adjoint Higgs field
into this setting. The addition of an adjoint scalar field to such
a theory allows us to examine the interplay of confinement and the
Higgs mechanism. For a quartic scalar field potential $V\left(\phi\right)$,
there is a $Z(2)_{H}$ global symmetry given by $\phi\rightarrow-\phi$.
Unlike the gauge coupling, the quartic interaction $\lambda$ of such
a scalar is not asymptotically free. However, we are free to set the
running coupling $\lambda\left(\mu\right)$ so that it is small at
the scale $\mu=1/L$, and semiclassical methods, including perturbation
theory, are valid. An $SU(N)$ adjoint scalar Higgs model on $R^{3}\times S^{1}$
has a natural global symmetry group $Z(N)_{C}\times Z(2)_{H}$. We
will focus in what follows on the case $N=2$. Not only is it the
simplest case, but for $N\ge3$, the gauge theory on $R^{3}\times S^{1}$
has additional phases intermediate between the confined and deconfined
phases, complicating the analysis \cite{Myers:2007vc}. 

The supersymmetric analog of this model is the Seiberg-Witten
model \cite{Seiberg:1994rs}, which is an $\mathcal{N}=2$ supersymmetric gauge theory
with gauge group $SU(2)$. Seiberg and Witten found that in this model
the addition of an $\mathcal{N}=1$ mass perturbation leads to confinement
by magnetic monopoles. Recently, Poppitz and Unsal have examined the
behavior of this model on $R^{3}\times S^{1}$, and concluded that
the confined phase seen for small compactification circumference on
$R^{3}\times S^{1}$ is connected to the confining phase at infinite
compactification circumference. In their work, Euclidean monopoles
in which a linear combination of $A_{4}$ and $\phi$ plays the role
of the scalar field appear prominently, in a very similar fashion
to the nonsupersymmetric model \cite{Nishimura:2010xa}.

The scalar field $\phi$ is not gauge invariant, and cannot serve
as an order parameter for the breaking of the $Z(2)_{H}$ symmetry
associated with $\phi$ when gauge interactions are present. This
is an old problem, a consequence of Elitzur's theorem \cite{Elitzur:1975im}.
Higgs models with scalar fields in the fundamental and adjoint representations
behave differently. For Higgs models with scalar fields in the fundamental
representation, the confined and Higgs phases are connected \cite{Fradkin:1978dv}
in a manner similar to the connection between liquid and gas phases.
In this case, the $Z(N)_{C}$ center symmetry is explicitly broken,
and large Wilson loops do not have area-law behavior due to screening
by the scalars.
In the adjoint case, $Z(N)_{C}$ center symmetry is preserved by the
action, and there is a distinct phase transition between the 
confined and Higgs phases.
In the $R^{3}\times S^{1}$ model we consider, we will show that there
are combinations of $\phi$ and $P$, such as $Tr_{F}\phi P$
that can serve as gauge-invariant order parameters for the symmetries
of the model.

Section II describes in detail the effective potential for the Polyakov
loop and deformations added to it that restore confinement at small
$L$, focusing on the case of $SU(2)$. We will discuss a number of
possible deformation terms and their effect on the order of the deconfining
phase transition. We will show that a particularly useful deformation
can be obtained by considering the embedding of two-dimensional fermions
into the four-dimensional theory. This deformation leads to a
simple treatment of the perturbatively determined phase diagram in
Sec.\ III, although our overall conclusions regarding the phase
structure are general. In Sec.~III we determine the phase structure
of the $SU(2)$ model using the effective potential $U_{eff}$, evaluated
at one loop. The evaluation at finite $L$ of the functional determinants
representing one-loop contributions to $U_{eff}$ will be the same
as those needed at finite temperature. With the inclusion of a deformation
term, we will show that there are four different phases in perturbation
theory, corresponding to different patterns of symmetry breaking:
a deconfined phase, a confined phase, a Higgs phase, and a phase which
appears to exhibit both the Higgs mechanism and confinement. 
In Sec.~IV we show that the phase that apparently combines confinement and
the Higgs mechanism is in fact a mixed confined phase, where the $Z(2)_{C}\times Z(2)_{H}$
global symmetry breaks spontaneously to the $Z(2)$ subgroup that
acts nontrivially on both the scalar field and the Polyakov loop.
We show that three gauge-invariant order parameters $Tr_{F}P$, $Tr_{F}\phi P$
and $Tr_{F}\phi P^{2}$ are sufficient to resolve the phase structure,
and characterize all four phases in terms of their global symmetries.
Section V finds in the BPS limit the classical solutions of the Euclidean
equations of motion that appear as constituents of instantons in the
pure gauge case. These solutions are not identical to Minkowski-space
monopoles, which also occur in this model. The most interesting and
general case is the mixed confined phase, where both $\phi$ and $A_{4}$
have expected values in an appropriately chosen gauge. In the mixed
confined phase, a linear combination of $\phi$ and $A_{4}$ plays
the same role that $A_{4}$ plays in the analysis of pure gauge theories
on $R^{3}\times S^{1}$, in line with the breaking $Z(2)_{C}\times Z(2)_{H}\rightarrow Z(2)$.
The behavior of the monopole solutions in the other three phases appear
as special cases of the mixed confined phase. Section VI discusses
the effects of these Euclidean monopoles on the dynamics of the model.
In previous studies of the confined phase in $SU(N)$ gauge theories
on $R^{3}\times S^{1}$, it has been shown that Euclidean-space monopoles
play a key role in the area-law behavior of Wilson loops in planes
orthogonal to the compact direction \cite{Unsal:2007vu,Unsal:2008ch,Unsal:2007jx}. 
It is therefore no surprise to
find that monopoles play an important role when an adjoint scalar field
is present. However, there is great subtlety and variety in the analysis.
Nevertheless, we show that a dilute monopole gas gives rise to confining
behavior for Wilson loops over a broad region that includes part of
all four phases. A final section gives our conclusions.

\section{Role of the deformation}

As explained in the introduction, the one-loop gauge boson effective
potential $V_{g}$ favors the deconfined phase. In the case of $SU(2)$,
where the Polyakov loop can be parametrized as $Tr_{F}P=2\cos(\theta)$,
$V_{g}$ can be written as \cite{Gross:1980br,Weiss:1980rj,Weiss:1981ev,Meisinger:2001fi}
%\begin{eqnarray}
%V_{g} & = & -\frac{2}{\pi^{2}L^{4}}\sum_{n=1}^{\infty}\frac{Tr_{A}P^{n}}{n^{4}}\nonumber\\
%& = & -\frac{\pi^{2}}{15L^{4}}+\frac{4}{3\pi^{2}L^{4}}\theta^{2}\left(\theta-\pi\right)^{2}
%\end{eqnarray}
\begin{equation}
V_{g}  =  -\frac{2}{\pi^{2}L^{4}}\sum_{n=1}^{\infty}\frac{Tr_{A}P^{n}}{n^{4}}
\end{equation}
or equivalently
\begin{equation}
V_{g} =  -\frac{\pi^{2}}{15L^{4}}+\frac{4}{3\pi^{2}L^{4}}\theta^{2}\left(\theta-\pi\right)^{2}
\end{equation}
which is minimized at $\theta=0$ or $\pi$ corresponding to $Tr_{F}P=\pm2$.
In order to realize the confined phase for small $L$, we will add
a double-trace deformation term $S_{d}$ to the action. This term
will be a $Z(N)_{C}$-invariant function of $P$, and therefore will
be nonlocal in the compact variable $x_{4}.$ Many forms of $S_{d}$
may be used, such that the confined phase is favored for some range
of parameters. In the case of $SU(2),$ there is an interesting issue
concerning the order of the transition from the confined to the deconfined
phase. Although the pure gauge theory is clearly related to three-dimensional
$Z(2)$ spin systems, this does not ensure a second-order transition
in the Ising model universality class because the transition may be
first order. This issue can easily be understood from the point of
view of Landau-Ginzburg theory. Consider a general theory with a real
scalar order parameter $\rho$ and a Landau-Ginzburg free energy density
$f[\rho]$ which is $Z(2)$ invariant. It may be expanded as
\begin{equation}
f\left[\rho\right]=\frac{1}{2}r\rho^{2}+\frac{1}{4!}\lambda\rho^{4}+\frac{1}{6!}\kappa\rho^{6}
\end{equation}
with $\kappa>0$ for global stability. As long
as $\lambda>0$, the transition will be second-order, occurring at
$r=0$. However, if $\lambda<0$, it is easy to see that the transition
may be first-order \cite{cha95}. Because we have a high
degree of freedom in choosing our deformation in $SU(2)$, we can
also choose the order of the transition.

\begin{figure}
\includegraphics[width=5.5in]{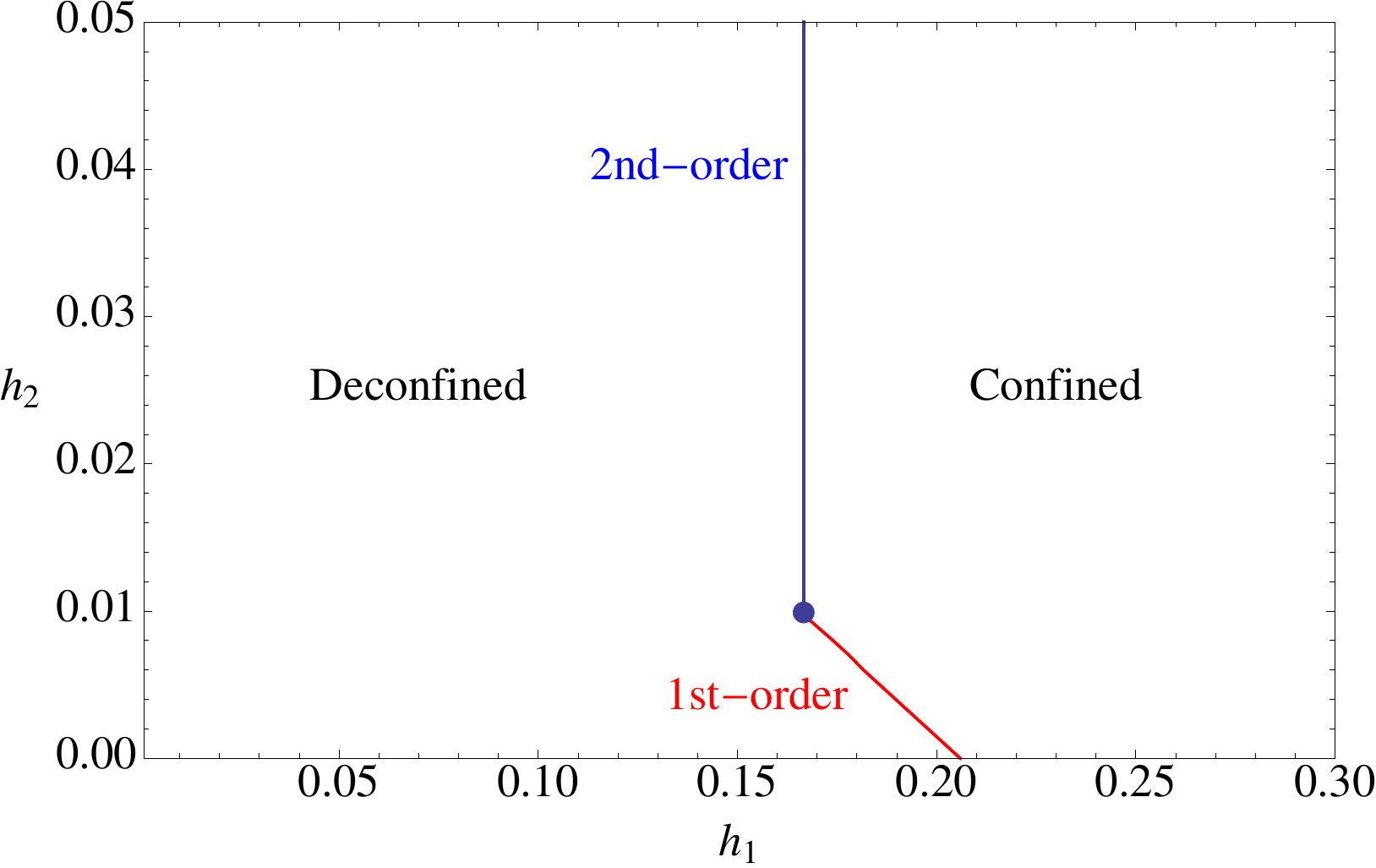}\caption{\label{fig:h1-h2}Phase diagram of an $SU(2)$ gauge theory as a 
function of the $h_{1}$ and $h_{2}$ deformation parameters.}
\end{figure}

A minimal choice for $S_{d}$ takes the form
\begin{equation}
S_{d}=L\int d^{3}x\,\frac{h_{1}}{L^{4}}\left|Tr_{F}P\left(\vec{x}\right)\right|^{2}
\end{equation}
which favors the confined phase with $Tr_{F}P=0$ for $h_{1}>0$.
From a Landau-Ginzburg point of view, changing $h_{1}$ is a change
to $r$. However, the transition between the confined and deconfined
phases is first-order when
$S_{d}$ is added to the one-loop effective action of the gauge theory. 
It is instructive to consider a slightly generalized form
\begin{equation}
V_{d}=h_{1}L^{-4}\left(Tr_{F}P\right)^{2}+h_{2}L^{-4}\left(Tr_{F}P\right)^{4}
\end{equation}
where we define the potential $V_{d}$ via
\begin{equation}
S_{d}=L\int d^{3}x\, V_{d}.
\end{equation}
For sufficiently large $h_{1}>0$, the symmetry will be restored.
Expanding the gluon potential with this deformed potential around
the symmetric point, $\theta=\pi/2$, we get a potential of Landau-Ginzburg
type
\begin{eqnarray}
V_{g}+V_{d} & \simeq & \frac{\pi^{2}}{60L^{4}}+\frac{4}{L^{4}}\left(h_{1}-\frac{1}{6}\right)\left(\theta-\frac{\pi}{2}\right)^{2}+\frac{4}{3L^{4}}\left(-h_{1}+12h_{2}+\frac{1}{\pi^{2}}\right)\left(\theta-\frac{\pi}{2}\right)^{4} \nonumber \\
 &  & +\frac{8}{45L^{4}}\left(h_{1}-60h_{2}\right)\left(\theta-\frac{\pi}{2}\right)^{6}
\end{eqnarray}
displaying explicitly the variation of the low-order terms in the
expansion. If the phase transition is second-order, it must occur
at $h_{1}=1/6$. However, if the coefficient of the quartic term is
negative, the confined phase at $\theta=\pi/2$ will be unstable when
$h_{1}=1/6$. This tells us that the transition is first-order for
sufficiently small $h_{2}.$ On the other hand, when $h_{2}$ is large,
we can ignore terms past quartic because $\theta$ is bounded, and
the transition is second-order. The tricritical point where the transition
changes from first- to second-order, lies somewhere on the line of
$h_{1}=1/6,$ but it must be located numerically. 
We plot the phase diagram of the deformed $SU(2)$ as
shown in Fig.~\ref{fig:h1-h2}. 

Another possibility is to choose a form for $V_{d}$ which is proportional
to the one-loop expression for the gauge boson contribution to the
effective action, but with opposite sign \cite{Myers:2007vc}: 
\begin{equation}
V_{d}=\frac{2h}{\pi^{2}L^{4}}\sum_{n=1}^{\infty}\frac{\left|Tr_{F}P^{n}\right|^{2}}{n^{4}}.
\end{equation}
The action will cancel the leading-order $1/L^{4}$ contribution of
the gauge bosons to the effective action when $h=1$, and the confined
phase will be favored at small $L$ when $h>1$. This choice for $V_{d}$
leads to a very strong first-order transition as $h$ is varied between
a confined phase where $Tr_{F}P$=0 and a deconfined phase where $Tr_{F}P=\pm2$
, the largest possible value. This form for the deformation can be
approximately implemented by a local addition to the action, corresponding
to $N_{f}$ flavors of adjoint Dirac fermions of mass $M$ with periodic
boundary conditions in the compact direction. In general, the potential
for such adjoint fermions in $\left(d+1\right)$-dimension is
\begin{equation}
\label{1lF}
4N_{f}s\left(\frac{M}{2\pi L}\right)^{\left(d+1\right)/2}\sum_{n=1}^{\infty}\frac{K_{\left(d+1\right)/2}\left(nML\right)Tr_{A}P^{n}}{n^{\left(d+1\right)/2}}
\end{equation}
where $K_{(d+1)/2}$ is the modified Bessel function and $s$ accounts
for spin degeneracy \cite{Meisinger:2001fi}. In the limit $M\rightarrow0$
with $d=3$ spatial dimensions, the adjoint fermions will make a one-loop
contribution to the effective potential of the form given above with
the identification $h=sN_{f}$, up to a term independent of $Tr_{F}P$
because $Tr_{A}P^{n}=\left|Tr_{F}P^{n}\right|^{2}-1$. The transition
between phases is first-order for all $M$.

\begin{figure}
\includegraphics[width=5.5in]{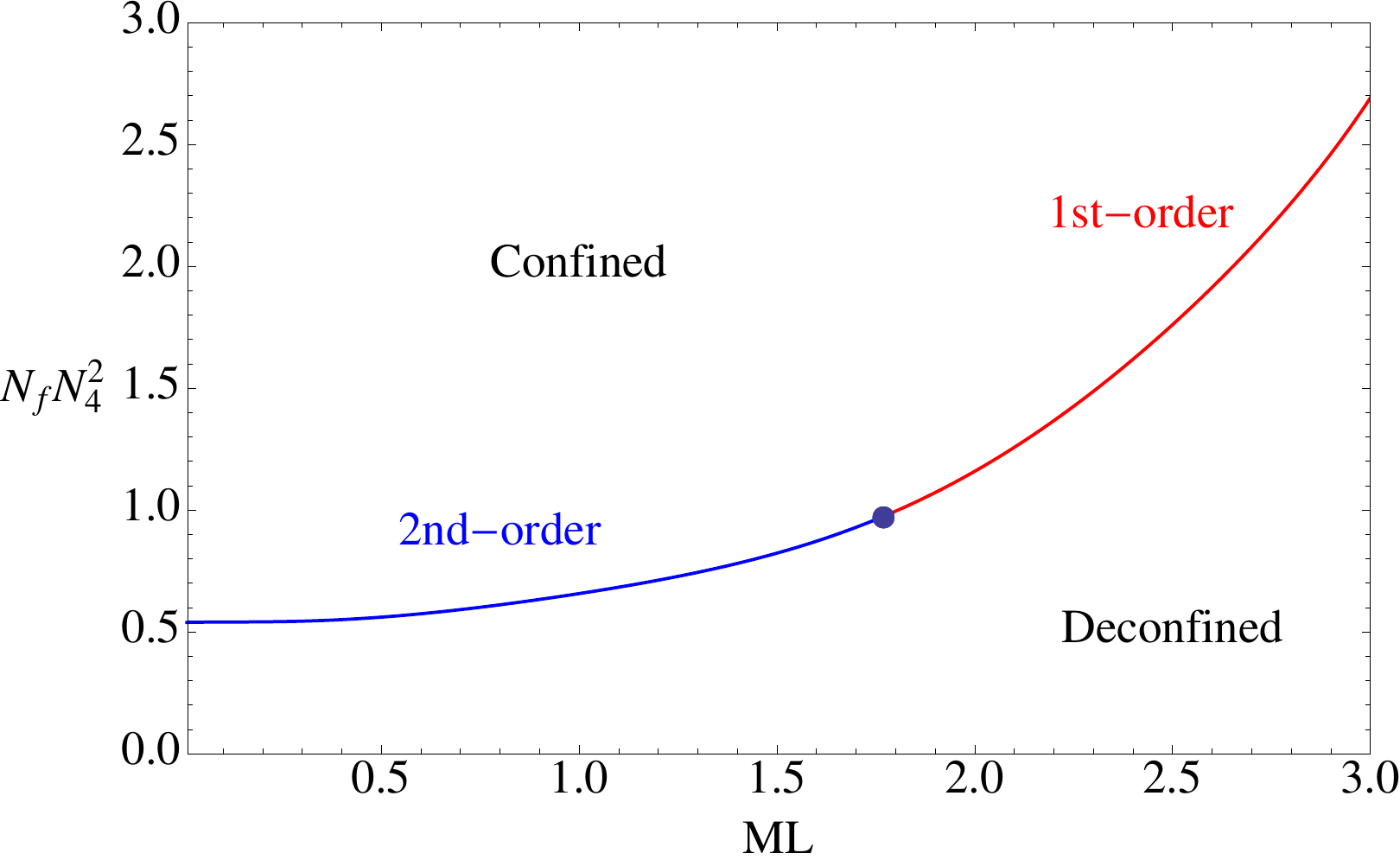}\caption{\label{fig:2dfermion}The phase diagram of an $SU(2)$ gauge theory
with a deformation inspired by $N_{f}$ two-dimensional fermions of
mass $M$ as a function of $ML$ and $N_{f}N_{4}^{2}$.}
\end{figure}

The most analytically tractable choice we have found that yields a
second-order transition is based on the one-loop potential for $N_{f}$
adjoint Dirac fermions with periodic boundary conditions in two dimensions
instead of four, i.e., $d=1$ in Eq.~\ref{1lF}, yielding in two dimensions
\begin{equation}
\frac{2MLN_{f}}{\pi L^{2}}\sum_{n=1}^{\infty}\frac{K_{1}\left(nML\right)Tr_{A}P^{n}}{n}.
\end{equation}
These sheets of two-dimensional fermions can be embedded in four dimensions
with a density $1/a^{2}$ in the plane orthogonal to the plane of
the fermions. Then $V_{d}$ is given by
\begin{equation}
V_{d}=\frac{2MLN_{f}}{\pi a^{2}L^{2}}\sum_{n=1}^{\infty}\frac{K_{1}\left(nML\right)Tr_{A}P^{n}}{n}.
\end{equation}
 In a lattice implementation, we would identify $ $$a$ as the lattice
spacing and an overall coefficient of order one would depend on the
lattice fermion implementation. Using the relation $L=N_{4}a$, where
$N_{4}$ is the number of lattice sites in the compact direction,
we would have
\begin{equation}
V_{d}=\frac{2MLN_{f}N_{4}^{2}}{\pi L^{4}}\sum_{n=1}^{\infty}\frac{K_{1}\left(nML\right)Tr_{A}P^{n}}{n}.
\end{equation}
The infinite series can be summed exactly in the limit when the mass
goes to zero,
\begin{eqnarray}
\lim_{M\rightarrow0}V_{d} & = & \frac{2N_{f}N_{4}^{2}}{\pi L^{4}}\sum_{n=1}^{\infty}\frac{Tr_{A}P^{n}}{n^{2}}=\frac{4N_{f}N_{4}^{2}}{\pi L^{4}}\left(\theta-\pi/2\right)^{2}
\end{eqnarray}
where $0\le\theta\le\pi$. This deformation leads to a second-order
phase transition at some $N_{f}$ for sufficiently small $M.$ We
stress that although the form of the deformation term was motivated
by the connection with adjoint fermions, it is in fact a deformation
term with no additional dynamical degrees of freedom. Because we treat
this term as a deformation, we can identify the compactification circumference
as an inverse temperature $L=\beta$; this would not be legitimate
for periodic adjoint fermions, because spectral positivity in the
compact direction would fail. We minimize the effective potential
of gluons with this deformation numerically by changing the two dimensionless
parameters, $N_{f}N_{4}^{2}$ and $ML$, and construct the phase diagram
as shown in Fig.~\ref{fig:2dfermion}. As we increase $ML$, the
contribution from adjoint fermions is suppressed, so a larger number
of flavors is needed to retain confinement. However, the transition
becomes first-order for sufficiently high $ML$ as we change $N_{f}N_{4}^{2}$.
The tricritical point lies on $\left(ML,N_{f}N_{4}^{2}\right)_{c}\simeq\left(1.771,0.955\right)$.
We will use the $M=0$ form in what follows, thereby obtaining a second-order
deconfinement transition. 

\section{The Effective Potential}

The phase diagram of our $SU(2)$ model will be calculated from an
approximate form of the one-loop effective potential, including the
deformation term. The effective potential will be calculated in background
field gauge \cite{Abbott:1980hw,Abbott:1981ke}, with the background fields consisting
of a scalar expectation value for $\phi$ and a constant value for $A_{4}$;
the latter gives rise to a nontrivial Polyakov loop background. For a general
Higgs theory, the classical Euclidean action can be written as
\begin{equation}
S_{c}=\int d^{4}x\left[\frac{1}{4}\left(F_{\mu\nu}^{a}\right)^{2}+\frac{1}{2}\left(D_{\mu}\phi\right)^{T}\cdot D_{\mu}\phi+V\left(\phi\right)\right]
\end{equation}
where the field $\phi$ is in an arbitrary real representation $R$
of the gauge group $G$ of dimension $n$, in general reducible. The
index $a$ runs over the number of generators of the group, $N^{2}-1$
for $SU(N)$.

The potential $V\left(\phi\right)$ we use is given by
\begin{equation}
V\left(\phi\right)=\frac{1}{2}m^{2}\phi^{2}+\frac{1}{4}\lambda\left(\phi^{2}\right)^{2}
\end{equation}
where $\phi^{2}=\phi^{T}\phi$. The covariant derivative acts on $\phi$
as
\begin{equation}
D_{\mu}\left(A\right)\phi=\partial_{\mu}\phi-igA_{\mu}\phi
\end{equation}
where the gauge field is written as an $n\times n$ matrix using $A_{\mu}=A_{\mu}^{a}T^{a}$
where the $T^{a}$ are the generators of the group in the representation
$R$. The field strength tensor in a matrix notation is 
\begin{equation}
F_{\mu\nu}=\partial_{\mu}A_{\nu}-\partial_{\nu}A_{\mu}-ig\left[A_{\mu},A_{\nu}\right]
\end{equation}
 or 
\begin{equation}
F_{\mu\nu}^{a}=\partial_{\mu}A_{\nu}^{a}-\partial_{\nu}A_{\mu}^{a}+gf^{abc}A_{\mu}^{b}A_{\nu}^{c}
\end{equation}
 in terms of components.

The classical contribution to the effective
potential is the sum of the scalar potential $V(\phi)$ and a contribution
from the kinetic term:
\begin{equation}
V_{c}(\phi)=-\frac{1}{2}g^{2}\left(A_{\mu}\phi\right)^{T}\cdot A_{\mu}\phi+V(\phi).
\end{equation}
The contribution from the kinetic term is positive-definite, despite
appearances. In a real representation, the Hermitian generators $T^{a}$
are purely imaginary, so $T^{a*}=-T^{a}$. This in turn implies $T^{aT}=-T^{a}$,
and thus $A_{\mu}^{T}=-A_{\mu}$. In the case of the adjoint representation,
this term can be written in matrix notation as
\begin{equation}
-g^{2}Tr_{F}\left[A_{4},\phi\right]^{2}
\end{equation}
where $\left[A_{4},\phi\right]$ is clearly anti-Hermitian. The positivity
of this term for the adjoint representation implies that the effective
potential will be minimized if $\left[A_{4},\phi\right]=0$.

The calculation of the effective potential in the presence of a background
Polyakov loop is similar to the case of finite temperature and density
\cite{Loewe:2005df}, because a chemical potential is an imaginary
$U(1)$ background $A_{4}$ expected value. The one-loop effective
action $\Gamma$ for the Higgs model without the deformation term
is given by the classical action $S$ plus contributions from the
functional determinants of the gauge, scalar and ghost fields. The
computation is only simple in $R_\xi$ gauge with $\xi=1$. For the
adjoint scalar model, the result is
\begin{equation}
\Gamma=S+Tr_{A}\log\left[\left(-\bar{D}_{\mu}^{2}\right)^{ac}+\left(M_{g}^{2}\right)^{ac}\right]+\frac{1}{2}Tr_{R}\log\left[\left(-\bar{D}_{\mu}^{2}\right)+M_{s}^{2}\right]
\end{equation}
where the functional traces are taken over space-time as well as the
internal symmetry group and $\bar{D}_{\mu}$ is the covariant derivative
with respect to the background field. We denote the background fields
by $\bar\phi$ and $\bar{A}_\mu$.
The first trace represents the
net contribution of the gauge and ghost fields, while the second term
is the contribution of the scalar field. The mass matrices depend
on the background field configuration and are given by
\begin{equation}
\left(M_{g}^{2}\right)^{ac}=g^{2}\bar{\phi}^{T}T^{a}T^{c}\bar{\phi}
\end{equation}
for the gauge fields and
\begin{equation}
M_{s}^{2}=m^{2}+\lambda\bar{\phi}^{2}+2\lambda\bar{\phi}\bar{\phi}^{T}+g^{2}T^{a}\bar{\phi}\bar{\phi}^{T}T^{a}
\end{equation}
 for the scalar fields. For static background fields we have
\begin{equation}
\Gamma=\int d^{4}x\, U_{eff}.
\end{equation}
The contribution to the effective potential from the functional determinants
may be separated into a contribution independent of $L$, analogous
to $T=0,$ of the form
\begin{equation}
V_{1l}^{\infty}=2\frac{1}{64\pi^{2}}Tr_{A}\left[\left(M_{g}^{2}\right)^{2}\log\left(M_{g}^{2}/\Lambda^{2}\right)\right]+\frac{1}{64\pi^{2}}Tr_{R}\left[\left(M_{s}^{2}\right)^{2}\log\left(M_{s}^{2}/\Lambda^{2}\right)\right]
\end{equation}
where the traces are taken over representations of the gauge group,
with $A$ denoting the adjoint representation. $\Lambda$ is the usual
scale-setting parameter with dimensions of mass required by renormalization.
There is also an $L$-dependent contribution, corresponding to $T\ne0$,
of the form $V_{1l}^{L}=V_{1lg}^{L}+V_{1l\phi}^{L}$ where 
\begin{equation}
V_{1lg}^{L}=\frac{2}{L}Tr_{A}\int\frac{d^{3}p}{\left(2\pi\right)^{3}}\log\left[1-P\:\exp\left(-L\sqrt{p^{2}+M_{g}^{2}}\right)\right]
\end{equation}
 and
\begin{equation}
V_{1l\phi}^{L}=\frac{1}{L}Tr_{R}\int\frac{d^{3}p}{\left(2\pi\right)^{3}}\log\left[1-P\:\exp\left(-L\sqrt{p^{2}+M_{s}^{2}}\right)\right]
\end{equation}
where $P$ is simply $\exp\left(igL\bar{A}_{4}\right)$. We have assumed
in these expressions that the mass matrices are diagonal, and so commute
with the Polyakov loop, as is the case if $\phi$ is in the adjoint
representation of $SU(N)$.

We now specialize to the case of adjoint $SU(2)$ where we take $\bar{\phi}=\left(0,0,v\right)$
and $P=diag\left[\exp\left(i\theta\right),\exp\left(-i\theta\right)\right]$
in the fundamental representation. It is easy to check that the gauge
boson mass matrix has the form
\begin{equation}
M_{g}^{2}=\left(\begin{array}{ccc}
g^{2}v^{2}\\
 & g^{2}v^{2}\\
 &  & 0
\end{array}\right)
\end{equation}
 and the scalar mass matrix $M_{s}^{2}$ is
\begin{equation}
M_{s}^{2}=\left(\begin{array}{ccc}
m^{2}+\lambda v^{2}+g^{2}v^{2}\\
 & m^{2}+\lambda v^{2}+g^{2}v^{2}\\
 &  & m^{2}+3\lambda v^{2}
\end{array}\right).
\end{equation}
 The complete one-loop effective potential for the scalar-gauge system
is then
\begin{eqnarray}
V_{eff} & = & \frac{1}{2}m^{2}v^{2}+\frac{1}{4}\lambda v^{4}+\frac{2\cdot2}{64\pi^{2}}g^{4}v^{4}\log\left(g^{2}v^{2}/\Lambda^{2}\right)+\frac{2}{64\pi^{2}}\left(m^{2}+\lambda v^{2}+g^{2}v^{2}\right)^{2}\log\left[\left(m^{2}+\lambda v^{2}+g^{2}v^{2}\right)/\Lambda^{2}\right] \nonumber \\
 &  & +\frac{1}{64\pi^{2}}\left(m^{2}+3\lambda v^{2}\right)^{2}\log\left[\left(m^{2}+3\lambda v^{2}\right)/\Lambda^{2}\right]+2\frac{1}{L}\int\frac{d^{3}p}{\left(2\pi\right)^{3}}\log\left[1-e^{2i\theta}e^{-L\sqrt{p^{2}+g^{2}v^{2}}}\right] \nonumber \\
 &  & +2\frac{1}{L}\int\frac{d^{3}p}{\left(2\pi\right)^{3}}\log\left[1-e^{-2i\theta}e^{-L\sqrt{p^{2}+g^{2}v^{2}}}\right]+2\frac{1}{L}\int\frac{d^{3}p}{\left(2\pi\right)^{3}}\log\left[1-e^{-L\left|p\right|}\right] \nonumber \\
 &  & +\frac{1}{L}\int\frac{d^{3}p}{\left(2\pi\right)^{3}}\log\left[1-e^{2i\theta}e^{-L\sqrt{p^{2}+m^{2}+\lambda v^{2}+g^{2}v^{2}}}\right] \nonumber \\
 &  & +\frac{1}{L}\int\frac{d^{3}p}{\left(2\pi\right)^{3}}\log\left[1-e^{-2i\theta}e^{-L\sqrt{p^{2}+m^{2}+\lambda v^{2}+g^{2}v^{2}}}\right] \nonumber \\
 &  & +\frac{1}{L}\int\frac{d^{3}p}{\left(2\pi\right)^{3}}\log\left[1-e^{-L\sqrt{p^{2}+m^{2}+3\lambda v^{2}}}\right].
\end{eqnarray}

As explained in the introduction, the extra term $S_{d}$ added to
the action is used to offset one-loop terms in $\Gamma$ that favor
the deconfined phase. These one-loop terms are $O(1)$ in the loop
expansion, whereas the classical action $S_{c}$ is $O(\hbar^{-1})$.
It is thus consistent to take $S_{d}$ to be $O(1)$ in the loop expansion.
This occurs naturally when the added term $S_{d}$ represents fermions
in the adjoint representation, but in the case of a deformation it
is essentially a choice we make in defining what our perturbation
theory is. The total one-loop effective potential is
\begin{equation}
U_{eff}=V_{eff}+V_{d}
\end{equation}
which is a function of the expected values $\bar{\phi}$ and $\bar{A}_{4}$
and depends on the parameters $g$, $m^{2}$, $\lambda$ and $L$,
as well as any additional parameters in $V_{d}$. We use the form
of $V_{d}$ given in Sec.~II:
\begin{equation}
V_{d}=\frac{4N_{f}N_{4}^{2}}{\pi L^{4}}\left(\theta-\pi/2\right)^{2}.
\end{equation}
We now make use of an approximate form for the integrals \cite{Meisinger:2001fi}
\begin{eqnarray}
V_{B} & = & \frac{1}{L}\int\frac{d^{3}p}{\left(2\pi\right)^{3}}\log\left[1-e^{i\theta}e^{-L\sqrt{p^{2}+M^{2}}}\right]+\frac{1}{L}\int\frac{d^{3}p}{\left(2\pi\right)^{3}}\log\left[1-e^{-i\theta}e^{-L\sqrt{p^{2}+M^{2}}}\right] \nonumber \\
 & \simeq & -\frac{2}{\pi^{2}L^{4}}\left[\frac{\pi^{4}}{90}-\frac{1}{48}\theta_{+}^{4}+\frac{\pi}{12}\theta_{+}^{3}-\frac{\pi^{2}}{12}\theta_{+}^{2}\right]+\frac{M^{2}}{2\pi^{2}L^{2}}\left[\frac{1}{4}\theta_{+}^{2}-\frac{\pi}{2}\theta_{+}+\frac{\pi^{2}}{6}\right] \nonumber \\
 &  & -\frac{M^{4}}{16\pi^{2}}\left[\ln\left(\frac{LM}{4\pi}\right)+\gamma-\frac{3}{4}\right]
\end{eqnarray}
where $\theta_{+}$ means $\theta$ made periodic over the range $0$
to $2\pi$ and $M$ is a mass term. Because we work with particles
in the adjoint representation, we must make the replacements $\theta_{+}\rightarrow2\theta$,
and the range of $\theta$ must be taken as $\left[0,\pi\right]$.
Applying this approximation to our complete expression for $V_{eff}$,
we have
\begin{eqnarray}
U_{eff} & = & \frac{1}{2}m^{2}v^{2}+\frac{1}{4}\lambda v^{4} \nonumber \\
 &  & -\frac{4}{\pi^{2}L^{4}}\left[\frac{\pi^{4}}{90}-\frac{1}{48}\left(2\theta\right)_{+}^{4}+\frac{\pi}{12}\left(2\theta\right)_{+}^{3}-\frac{\pi^{2}}{12}\left(2\theta\right)_{+}^{2}\right] \nonumber \\
 &  & +\frac{g^{2}v^{2}}{\pi^{2}L^{2}}\left[\frac{1}{4}\left(2\theta\right)_{+}^{2}-\frac{\pi}{2}\left(2\theta\right)_{+}+\frac{\pi^{2}}{6}\right]-\frac{g^{4}v^{4}}{16\pi^{2}}\left[\ln\left(\frac{L^{2}\Lambda^{2}}{16\pi^{2}}\right)+2\gamma-\frac{3}{2}\right] \nonumber \\
 &  & -\frac{2}{\pi^{2}L^{4}}\left[\frac{\pi^{4}}{90}\right]-\frac{2}{\pi^{2}L^{4}}\left[\frac{\pi^{4}}{90}-\frac{1}{48}\left(2\theta\right)_{+}^{4}+\frac{\pi}{12}\left(2\theta\right)_{+}^{3}-\frac{\pi^{2}}{12}\left(2\theta\right)_{+}^{2}\right] \nonumber \\
 &  & +\frac{\left(m^{2}+\lambda v^{2}+g^{2}v^{2}\right)}{2\pi^{2}L^{2}}\left[\frac{1}{4}\left(2\theta\right)_{+}^{2}-\frac{\pi}{2}\left(2\theta\right)_{+}+\frac{\pi^{2}}{6}\right] \nonumber \\
 &  & -\frac{\left(m^{2}+\lambda v^{2}+g^{2}v^{2}\right)^{2}}{32\pi^{2}}\left[\ln\left(\frac{L^{2}\Lambda^{2}}{16\pi^{2}}\right)+2\gamma-\frac{3}{2}\right] \nonumber \\
 &  & -\frac{1}{\pi^{2}L^{4}}\left[\frac{\pi^{4}}{90}\right]+\frac{m^{2}+3\lambda v^{2}}{4\pi^{2}L^{2}}\left[\frac{\pi^{2}}{6}\right]-\frac{\left(m^{2}+3\lambda v^{2}\right)^{2}}{64\pi^{2}}\left[\ln\left(\frac{L^{2}\Lambda^{2}}{16\pi^{2}}\right)+2\gamma-\frac{3}{2}\right]\nonumber \\
 &  & +\frac{4N_{f}N_{4}^{2}}{\pi L^{4}}\left(\theta-\pi/2\right)^{2}.
\end{eqnarray}
Note that the logarithmic dependence on the mass matrix disappears
in this small-$L$ expansion. Rearranging the leading-order terms,
we have
\begin{eqnarray}
U_{eff} & = & \frac{1}{2}m^{2}v^{2}+\frac{1}{4}\lambda v^{4} \nonumber \\
 &  & -\frac{6}{\pi^{2}L^{4}}\left[\frac{\pi^{4}}{90}-\frac{1}{48}\left(2\theta\right)_{+}^{4}+\frac{\pi}{12}\left(2\theta\right)_{+}^{3}-\frac{\pi^{2}}{12}\left(2\theta\right)_{+}^{2}\right]-\frac{3}{\pi^{2}L^{4}}\left[\frac{\pi^{4}}{90}\right] \nonumber \\
 &  & +\frac{\left(m^{2}+\lambda v^{2}+g^{2}v^{2}\right)}{2\pi^{2}L^{2}}\left[\frac{1}{4}\left(2\theta\right)_{+}^{2}-\frac{\pi}{2}\left(2\theta\right)_{+}+\frac{\pi^{2}}{6}\right] \nonumber \\
 &  & +\frac{2g^{2}v^{2}}{2\pi^{2}L^{2}}\left[\frac{1}{4}\left(2\theta\right)_{+}^{2}-\frac{\pi}{2}\left(2\theta\right)_{+}+\frac{\pi^{2}}{6}\right]+\frac{m^{2}+3\lambda v^{2}}{4\pi^{2}L^{2}}\left[\frac{\pi^{2}}{6}\right] \nonumber \\
 &  & -\frac{\left(m^{2}+3\lambda v^{2}\right)^{2}}{64\pi^{2}}\left[\ln\left(\frac{L^{2}\Lambda^{2}}{16\pi^{2}}\right)+2\gamma-\frac{3}{2}\right]-\frac{2g^{4}v^{4}}{32\pi^{2}}\left[\ln\left(\frac{L^{2}\Lambda^{2}}{16\pi^{2}}\right)+2\gamma-\frac{3}{2}\right] \nonumber \\
 &  & -\frac{\left(m^{2}+\lambda v^{2}+g^{2}v^{2}\right)^{2}}{32\pi^{2}}\left[\ln\left(\frac{L^{2}\Lambda^{2}}{16\pi^{2}}\right)+2\gamma-\frac{3}{2}\right]+\frac{4N_{f}N_{4}^{2}}{\pi L^{4}}\left(\theta-\pi/2\right)^{2}.
\end{eqnarray}
We now drop all the terms independent of $v$ and $\theta$ from $U_{eff}$.
Additionally, we define running couplings $m^{2}\left(L\right)$ and
$\lambda(L)$ in such a way that all one-loop contributions are included
in the running couplings when $\theta=\pi/2$
\begin{eqnarray}
U_{eff} & = & \frac{1}{2}m^{2}\left(L\right)v^{2}+\frac{1}{4}\lambda\left(L\right)v^{4} \nonumber \\
 &  & +\frac{1}{\pi^{2}L^{4}}\left[2\left(\theta-\frac{\pi}{2}\right)^{4}-\pi^{2}\left(\theta-\frac{\pi}{2}\right)^{2}\right]+\frac{4N_{f}N_{4}^{2}}{\pi L^{4}}\left(\theta-\pi/2\right)^{2} \nonumber \\
 &  & +\frac{\left(m^{2}+\lambda v^{2}+g^{2}v^{2}\right)}{2\pi^{2}L^{2}}\left(\theta-\pi/2\right)^{2}+\frac{2g^{2}v^{2}}{2\pi^{2}L^{2}}\left(\theta-\pi/2\right)^{2}.
\end{eqnarray}
In order for us to take the phase diagram predicted by our one-loop effective
potential seriously, both the gauge coupling $g(L)$ and the scalar
coupling $\lambda(L)$ must be small. The gauge coupling is naturally
small at a scale where $\Lambda L\ll1$ as a consequence of asymptotic
freedom, but the scalar coupling must be tuned to make $\lambda(L)$
small.

Naively, the phase diagram is controlled in perturbation theory by the two
quadratic terms
\begin{equation}
\frac{1}{2}m^{2}\left(L\right)v^{2}
\end{equation}
and 
\begin{equation}
\frac{1}{L^{4}}\left[\frac{4N_{f}N_{4}^{2}}{\pi}-1\right]\left(\theta-\frac{\pi}{2}\right)^{2}.
\end{equation}
 The potential also has a quartic coupling that couples together the two order parameters
in a way that generally can produce either four second-order transition lines meeting
at a tetracritical point or two second-order lines and one first-order line meeting at a
bicritical point \cite{cha95}. In the case at hand, the tetracritical phase diagram is obtained,
as we now show.
We define the parameter
\begin{equation}
a\equiv\frac{4N_{f}N_{4}^{2}}{\pi}-1.
\end{equation}
It is easy to see that there are at least two second-order phase transition lines that
meet at $\left( a=0 , m^2(L)=0 \right)$: one line is along $a=0$ for $m^2(L)>0$,
and the other is along $m^2(L)=0$ for $a>0$. Note that when $m^2(L)=0$,
the Lagrangian parameter $m^2$ is negative and $\mathcal{O}\left( \lambda/L^2 , g^2/L^2 \right)$.
It is easy to see that the critical line for $\theta$ is determined by the
$\mathcal{O}(1/L^4)$ terms in $U_{eff}$, 
implying the critical line is given by $a=0$ up to a term which is of order
$m^{2}(L)L^2$, which  is of order $\lambda$ or $g^2$ or less in the vicinity of
the tetracritical point. Thus to leading order in perturbation, the critical line
associated with $\theta$ is given by $a=0$. As $a$ moves from $a=0$ to negative
values, $\theta$ decreases from $\pi/2$, reaching $\theta=0$ at $a=-1$.
A given value of $a$ will determine the value of $\theta$, which in turn determines
the coefficient of a contribution to $U_{eff}$ of the form
\begin{equation}
\frac{\left( \lambda +3 g^{2}\right)}{2\pi^{2}L^{2}}\left(\theta-\pi/2\right)^{2}v^{2}.
\end{equation}
We can absorb this contribution into our definition of $m^2(L)$. This has
the effect of straightening out what would have been a curved segment in the critical line
associated with $\phi$ in the region $-1<a<0$; the critical line is straight in any
case for $a>0$, where $\theta=\pi/2$, and for $a<-1$, where $\theta=0$.
Henceforth, we will
write $m^{2}(L)$ as simply $m^{2}$ for notational simplicity.

We now see that the one-loop effective potential predicts two second-order phase transitions.
They appear to be essentially independent: when $m^{2}<0$, the scalar
expectation value $v$ is nonzero; for $m^{2}>0$, it is $0$. If
$a>0$, the angle $\theta$ associated with the Polyakov loop has
the value $\pi/2$, and $Z(2)_{C}$ center symmetry holds. For $a<0$,
center symmetry is broken. Thus there are four distinct phases. As
we have seen, the order of the deconfinement transition is nonuniversal,
depending on the deformation. The detailed structure of the phase
diagram will depend on the precise model. For example, if four-dimensional
adjoint fermions are used, a coupling of the form $Tr\left[\bar{\psi}\phi\psi\right]$
must be considered. A large value for $v$ gives rise to a large fermion
mass terms, which in turn reduces the ability of the adjoint fermions
to restore confinement \cite{Nishimura:2009me}. However, the basic
phase structure will be the same for all models. Three of the phases
are familiar: a Higgs phase, a confined phase and a deconfined phase.
However, the fourth phase, where center symmetry is unbroken and $v\ne0$
is novel, and appears to have some of the properties of both the confined
phase ($Tr_{F}P=0$) and the Higgs phase ($v\ne0$). In the next three
sections, we will explore this phenomenon, first in terms of symmetries
using the perturbative effective action, and then nonperturbatively.

\section{Symmetries and order parameters}

\begin{comment}
- consider renaming Higgs phase to Higgs/Polyakov phase
\end{comment}

An understanding of the overall phase structure can be based on the
global symmetries of this class of models. The action is invariant
under two global $Z(2)$ symmetries. The first symmetry, $Z(2)_{H}$
is the invariance of the action under a transformation of the scalar
field $\phi\rightarrow-\phi$. Because $\phi$ transforms under $SO(3)$,
the adjoint representation of $SU(2)$, this transformation is not
a gauge transformation, but a global symmetry. The other global symmetry,
$Z(2)_{C}$, is associated with the center symmetry of the $SU(2)$
gauge group, and is present because all fields have $0\, N$-ality.
Under this global symmetry, the action is invariant, but the Polyakov
loop $P$ transforms as $P\rightarrow-P$. It is useful to consider
three distinct gauge-invariant order parameters associated with the
$Z(2)_{C}\times Z(2)_{H}$ symmetry. Although these order parameters
are nonlocal in the compact direction, they are local in the three
noncompact directions. The first of these is the trace in the fundamental
representation of the Polyakov loop $P$ itself, $\left\langle Tr_{F}P\left(x\right)\right\rangle $,
which is independent of $x_{4}$. It transforms nontrivially under
$Z(2)_{C}$ but is invariant under $Z(2)_{H}$. The second is $\left\langle Tr_{F}\left[P^{2}\left(x\right)\phi(x)\right]\right\rangle $
which is invariant under $Z(2)_{C}$, but transforms nontrivially
under $Z(2)_{H}$. Finally, there is $\left\langle Tr_{F}\left[P\left(x\right)\phi(x)\right]\right\rangle $,
which transforms nontrivially under both groups.

In the deconfined phase, there is spontaneous breaking of $Z(2)_{C}$,
indicated by $\left\langle Tr_{F}P\left(x\right)\right\rangle \ne0$.
The Higgs phase is associated with the spontaneous breaking of $Z(2)_{H}$,
indicated by $\left\langle Tr_{F}\left[P^{2}\left(x\right)\phi(x)\right]\right\rangle \ne0$.
It appears that five distinct phases might be possible: a confined
phase, where $Z(2)_{C}\times Z(2)_{H}$ is unbroken; a deconfined
phase, where $Z(2)_{C}$ is spontaneously broken but $Z(2)_{H}$ is
unbroken; a Higgs phase, where both $Z(2)_{C}$ and $Z(2)_{H}$ are
spontaneously broken; a phase where $Z(2)_{H}$ is broken but $Z(2)_{C}$ is
unbroken; and finally a phase where $Z(2)_{C}\times Z(2)_{H}$ spontaneously
breaks to $Z(2)$. This last phase is only invariant under a simultaneous
transformation of $P$ and $\phi$. We will refer to this phase as
the mixed confined phase. The mixed confined phase in some sense takes
the place of a phase where $Z(2)_{H}$ is broken but $Z(2)_{C}$ is
unbroken, which would be a phase where both the Higgs mechanism and
confinement hold.

The minimum of the perturbative effective potential is specified
by the expected values $\theta$ and $v$. They are not themselves
gauge-invariant, but they can be used reliably to calculate gauge-invariant
order parameters for small $L$. We have simply
\begin{eqnarray}
\left\langle Tr_{F}P\left(x\right)\right\rangle  & = & 2\cos\left(\theta\right)\\
\left\langle Tr_{F}\left[P^{2}\left(x\right)\phi(x)\right]\right\rangle  & = & 2iv\sin\left(2\theta\right)\\
\left\langle Tr_{F}\left[P\left(x\right)\phi(x)\right]\right\rangle  & = & 2iv\sin\left(\theta\right).
\end{eqnarray}
The second and third expectation values are imaginary, but can be
made real if desired by forming the appropriate Hermitian operator.
The key technical point is that $\left\langle Tr_{F}\left[P^{2}\left(x\right)\phi(x)\right]\right\rangle $
is a gauge-invariant proxy for $\phi$ as long as $\sin\left(2\theta\right)\ne0$.
This restriction implies that the case of maximal center symmetry
breaking where $\theta\rightarrow0$ or $\theta\rightarrow\pi$ must
be treated as a limiting case. Although one-loop perturbation theory
does indicate maximal center symmetry breaking at high temperatures,
lattice simulations suggest that such temperatures are not reached
until well beyond the deconfinement transition. We assume that in
each phase where center symmetry is broken there is a region where
it is not maximally broken. It is easy to check that for our choice
of deformation this is the case.

\begin{figure}
\includegraphics[width=5.5in]{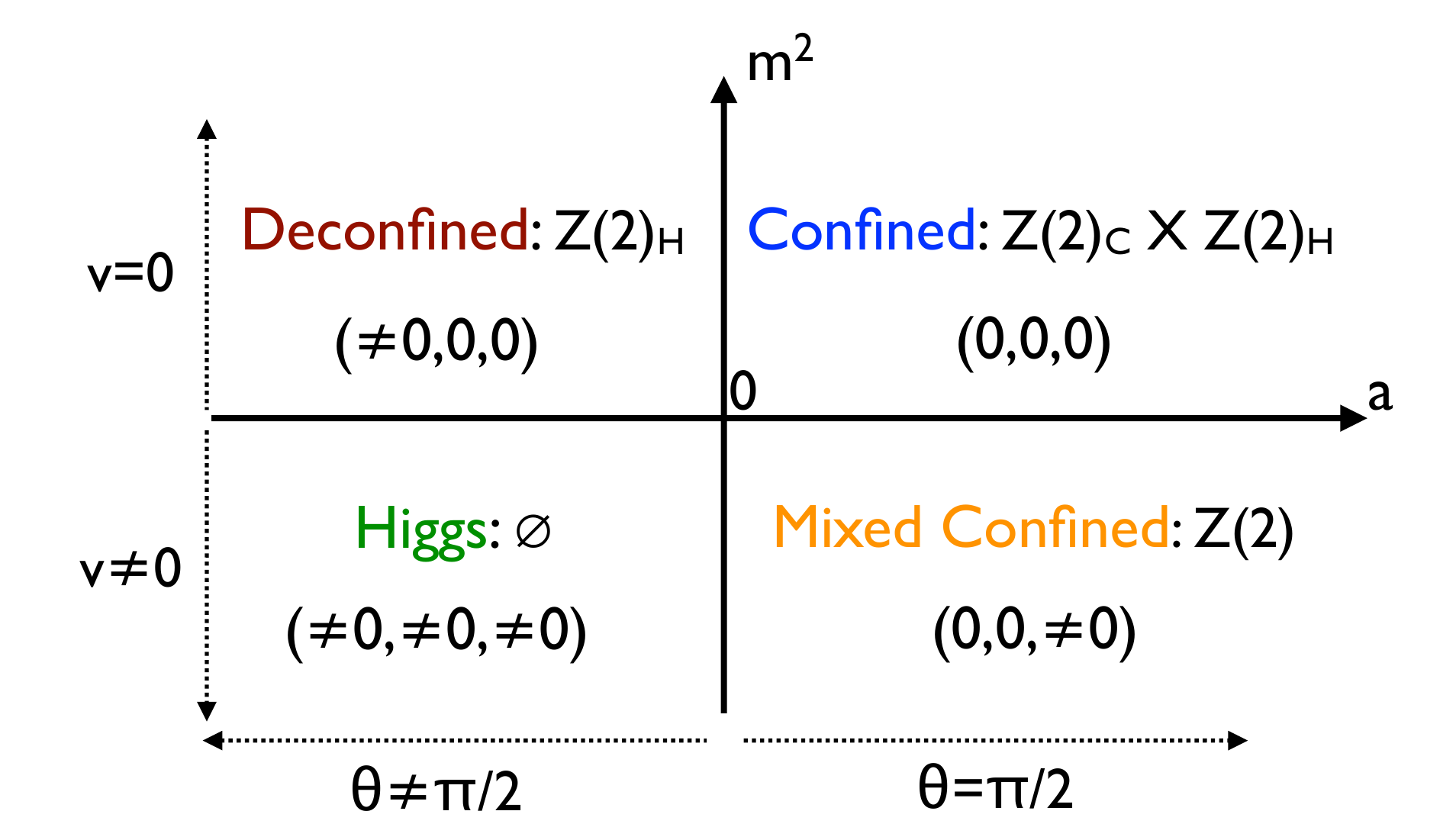}\caption{\label{fig:PhaseDiagramA} Phase diagram of $SU(2)$ Higgs model as
a function of $a$ and $m^{2}$. The values of the order parameters
are shown in parenthesis as $\left(\left\langle Tr_{F}P\right\rangle ,\left\langle Tr_{F}\left[P^{2}\phi\right]\right\rangle ,\left\langle Tr_{F}\left[P\phi\right]\right\rangle \right)$.}
\end{figure}

It is now easy to work out the phase diagram and the properties of
the phases, as shown in Fig.~\ref{fig:PhaseDiagramA} and Table
\ref{tab:Table1}. Naively, the phase that is both a confined and
a Higgs phase occurs when $a>0$ and $m^{2}<0$. This would be a phase
where $Z(2)_{H}$ is broken but $Z(2)_{C}$ is unbroken, in the sense
that $\left\langle Tr_{F}P\left(x\right)\right\rangle =0$ and $\left\langle Tr_{F}\left[P\left(x\right)\phi(x)\right]\right\rangle =0$
due to unbroken center symmetry, but $\left\langle Tr_{F}\left[P^{2}\left(x\right)\phi(x)\right]\right\rangle \ne0$
as in the Higgs phase. This behavior is not possible in perturbation
theory because $\left\langle Tr_{F}\left[P^{2}\left(x\right)\phi(x)\right]\right\rangle =0$
if $\left\langle Tr_{F}P\left(x\right)\right\rangle =0$. The phase
that replaces it is a confining phase because the Polyakov loop is
zero, but center symmetry has become entwined with the global symmetry
of the Higgs field. We will show in the next section that the nonperturbative
dynamics of the model shows the effects of this mixing in a direct
and dramatic way.

\begin{comment}
\begin{tabular}{|c|c|c|c|}
\hline 
Phase & $\left\langle Tr_{F}P\right\rangle $ & $\left\langle Tr_{F}\left[P^{2}\phi\right]\right\rangle $ & $\left\langle Tr_{F}\left[P\phi\right]\right\rangle $\tabularnewline
\hline 
\hline 
Confined & $0$ & $0$ & $0$\tabularnewline
\hline 
Deconfined & $\neq0$ & $0$ & $0$\tabularnewline
\hline 
Higgs & $\neq0$ & $\neq0$ & $\neq0$\tabularnewline
\hline 
Confined \& Higgs & $0$ & $\neq0$ & $0$\tabularnewline
\hline 
Mixed Confined & $0$ & $0$ & $\neq0$\tabularnewline
\hline 
\end{tabular}

\begin{tabular}{|c|c|c|c|c|}
\hline 
Phase & $\left\langle Tr_{F}P\right\rangle $ & $\left\langle Tr_{F}\left[P^{2}\phi\right]\right\rangle $ & $\left\langle Tr_{F}\left[P\phi\right]\right\rangle $ & Residual Symmetry\tabularnewline
\hline 
\hline 
Confined & $0$ & $0$ & $0$ & $Z(2)_{C}\times Z(2)_{H}$\tabularnewline
\hline 
Deconfined & $\neq0$ & $0$ & $0$ & $Z(2)_{H}$\tabularnewline
\hline 
Higgs & $\neq0$ & $\neq0$ & $\neq0$ & $\emptyset$\tabularnewline
\hline 
Confined \& Higgs & $0$ & $\neq0$ & $0$ & $Z(2)_{C}$\tabularnewline
\hline 
Mixed Confined & $0$ & $0$ & $\neq0$ & $Z(2)$\tabularnewline
\hline 
\end{tabular}
\end{comment}

\begin{table}
\begin{tabular}{|c|c|c|c|c|c|}
\hline 
Parameters & $\left\langle Tr_{F}P\right\rangle $ & $\left\langle Tr_{F}\left[P^{2}\phi\right]\right\rangle $ & $\left\langle Tr_{F}\left[P\phi\right]\right\rangle $ & Phase & Residual Symmetry\tabularnewline
\hline 
\hline 
$a>0$, $m^{2}>0$ & $0$ & $0$ & $0$ & Confined & $Z(2)_{C}\times Z(2)_{H}$\tabularnewline
\hline 
$a<0$, $m^{2}>0$ & $\neq0$ & $0$ & $0$ & Deconfined & $Z(2)_{H}$\tabularnewline
\hline 
$a<0$, $m^{2}<0$ & $\neq0$ & $\neq0$ & $\neq0$ & Higgs & $\emptyset$\tabularnewline
\hline 
$a>0$, $m^{2}<0$ & $0$ & $0$ & $\neq0$ & Mixed Confined & $Z(2)$\tabularnewline
\hline 
Absent & $0$ & $\neq0$ & $0$ & Confined \& Higgs & $Z(2)_{C}$\tabularnewline
\hline 
\end{tabular}

\caption{\label{tab:Table1}Properties of the four possible phases, along with
the Confined \& Higgs phase, which does not occur.}
\end{table}

\section{Classical Monopole Solutions}

The nonperturbative dynamics of gauge theories on $R^{3}\times S^{1}$
are all based on Polyakov's analysis of the Georgi-Glashow model in
three dimensions \cite{Polyakov:1976fu}. 
This is an $SU(2)$ gauge model coupled to an adjoint
Higgs scalar. The model we are considering thus differs by the addition
of a fourth compact dimension and a suitable deformation added to
the action. The four-dimensional Georgi-Glashow model is the standard
example of a gauge theory with classical monopole solutions when the
Higgs expectation value is nonzero. They are topologically stable
because $\Pi_{2}\left(SU\left(2\right)/U\left(1\right)\right)=\Pi_{1}\left(U\left(1\right)\right)=\mathbb{Z}$,
and make a nonperturbative contribution to the partition function
$Z$. In three dimensions, these monopoles are instantons. Polyakov
showed that a gas of such three-dimensional monopoles gives rise to
nonperturbative confinement in three dimensions, even though the
theory appears to be in a Higgs phase perturbatively. 

Because $L$ is small in our four-dimensional theory, the three-dimensional
effective theory describing the behavior of Wilson loops in the noncompact
directions will have many features in common with the three-dimensional
theory first discussed by Polyakov. In the four-dimensional theory,
monopole solutions with short worldline trajectories in the compact
direction exist, and behave as three-dimensional instantons in the
effective theory. It is useful to recall the analysis of the small-$L$
confined phase in the case of a gauge theory without scalars \cite{Myers:2007vc,Unsal:2008ch}.
In this theory, the role of the three-dimensional scalar field is
played by the fourth component of the gauge field $A_{4}$, which
has a vacuum expected value induced by the perturbative effective
potential. However, there is another way to understand the presence
of monopoles in this phase, based on studies of instantons in pure
gauge theories at finite temperature \cite{Lee:1998bb,Kraan:1998kp,Kraan:1998pm}.
If the Polyakov loop has a nontrivial expectation value, finite-temperature
instantons in $SU(N)$ may be decomposed into $N$ monopoles, and
the locations of the monopoles become parameters of the moduli space
of the instanton. In the case of $SU(2)$, an instanton may be decomposed
into a conventional BPS monopole 
and a so-called KK (Kaluza-Klein) monopole.
The presence of the KK monopole solution differentiates
the case of a gauge field at finite temperature from the case of an
adjoint scalar breaking $SU(N)$ to $U(1)^{N-1}$, in which case there
are $N-1$ fundamental monopoles. 

If a scalar field is added to the model, the coupling of $A_{4}$
to the $R^{3}$ gauge field $\vec{A}$ is identical to the coupling
of $\phi$ to $\vec{A}$, and nonzero expectation values for either
or both lead to topologically nontrivial field configurations. For
simplicity, we will continue to refer to these solutions as monopoles,
although they are instantons, in the sense that they are solutions
of the Euclidean field equations, and generally dyons in the sense
that $A_{4}$ has nontrivial behavior. In the general case, both
$A_{4}$ and $\phi$ play roles in the monopole solutions. This behavior
is similar to that found in Higgs models with more than one scalar
\cite{Weinberg:2006rq}. However, there is a significant difference.
When an adjoint Higgs model spontaneously breaks $SU(N)$ down to
$U(1)^{N-1}$, there are $N-1$ fundamental monopoles. When $A_{4}$
is responsible for the breaking of $SU(N)$ down to $U(1)^{N-1}$,
there is an additional monopole for a total of $N$ fundamental monopoles.
The solutions for all these monopoles can be found explicitly in the
BPS limit; when $A_{4}$ is nontrivial, the $N-1$ BPS monopoles
are joined by a KK  monopole \cite{Lee:1998bb,Kraan:1998kp,Kraan:1998pm}.
In what follows, it will be useful to differentiate between solutions
which saturate the Bogomolny bounds, versus solutions with the same
topological properties and reduce to the solutions saturating the
Bogomolny bounds in an appropriate limit. Thus we will distinguish
between BPS solutions and monopoles of BPS type, meaning monopoles
that reduce to BPS solutions in the appropriate limit. We will similarly
distinguish between KK monopoles and KK monopole solutions.

We will now show how the monopole solutions in the general case are
found. The monopole solutions in each of the four phases may be obtained
as special cases. We begin with the BPS-type solution where all fields
are independent of $x_{4}$. This construction is very similar to
the case of models with two Higgs fields \cite{Weinberg:2006rq}.
The Euclidean Lagrangian $\mathcal{{L}}$ is given by
\begin{equation}
\mathcal{{L}}  =  \frac{1}{4}\left(F_{\mu\nu}\right)^{2}+\frac{1}{2}\left(D_{\mu}\phi\right)^{2}+U_{eff}\left(\phi,A_{4}\right)
\end{equation}
which includes potential term for both $\phi$ and $A_{4}$. We assume
that $A_{4}$ commutes with $\phi$ so that $\mathcal{{L}}$ may be
reduced to
\begin{equation}
\mathcal{{L}}=\frac{1}{2}\left(D_{j}A_{4}\right)^{2}+\frac{1}{2}\left(B_{j}\right)^{2}+\frac{1}{2}\left(D_{j}\phi\right)^{2}+U_{eff}\left(\phi,A_{4}\right).
\end{equation}
We can associate with $\mathcal{{L}}$ an energy defined by
\begin{equation}
E=\int d^{3}x\left[\frac{1}{2}\left(B_{j}\right)^{2}+\frac{1}{2}\left(D_{j}A_{4}\right)^{2}+\frac{1}{2}\left(D_{j}\phi\right)^{2}+U_{eff}\left(\phi,A_{4}\right)\right]
\end{equation}
 as well as an action $S=LE$. We will concern ourselves for now with
the solutions in the BPS limit, in which the effective potential $U_{eff}$
is neglected, but the boundary conditions on $\phi$ and $A_{4}$
at infinity imposed by the potential are retained. 

We introduce two new fields
\begin{eqnarray}
b & = & \cos\alpha A_{4}+\sin\alpha\phi\\
c & = & -\sin\alpha A_{4}+\cos\alpha\phi
\end{eqnarray}
which are orthogonal linear combinations of $\phi$ and $A_{4}$,
depending on an arbitrary angle $\alpha$. We can write the energy
as
\begin{eqnarray}
E & = & \int d^{3}x\left[\frac{1}{2}B_{j}^{2}+\frac{1}{2}\left(D_{j}b\right)^{2}+\frac{1}{2}\left(D_{j}c\right)^{2}\right] \nonumber \\
 & = & \int d^{3}x\left[\frac{1}{2}\left(B_{j}\pm D_{j}b\right)^{2}+\frac{1}{2}\left(D_{j}c\right)^{2}\mp B_{j}D_{j}b\right].
\end{eqnarray}
This expression is a sum of squares plus a term which can be converted
to a surface integral, giving rise to the BPS inequality
\begin{equation}
E\geq\mp\int dS_{j}B_{j}b.
\end{equation}
The BPS inequality is saturated if the following equalities hold:
\begin{eqnarray}
B_{j} & = & \mp D_{j}b\nonumber\\
D_{j}c & = & 0.
\end{eqnarray}
For the case of a single monopole at the origin, we require the fields
at spatial infinity to behave as 
\begin{eqnarray}
\lim_{r\rightarrow\infty}\phi^{a} & = & v\frac{x^{a}}{r}\nonumber\\
\lim_{r\rightarrow\infty}A_{4}^{a} & = & w\frac{x^{a}}{r}\nonumber\\
\lim_{r\rightarrow\infty}A_{i}^{a} & = & \epsilon^{aij}\frac{x_{j}}{gr^{2}}.
\end{eqnarray}
Note that $w$ is related to the eigenvalues of $P$ at large distances
by $w=2\theta/gL$. The first two terms are the usual hedgehog fields.
$A_{i}^{a}$ is chosen such that covariant terms vanish at infinity:
$\left(D_{i}\phi\right)^{a}=0$ and $\left(D_{i}A_{4}\right)^{a}=0$.
With the 't Hooft-Polyakov ansatz, the general expressions for the
fields become 
\begin{eqnarray}
\phi^{a} & = & vf\left(r\right)\frac{x^{a}}{r}\nonumber\\
A_{4}^{a} & = & wh\left(r\right)\frac{x^{a}}{r}\nonumber\\
A_{i}^{a} & = & a\left(r\right)\epsilon^{aij}\frac{x_{j}}{gr^{2}}
\end{eqnarray}
where we define $v,\, w>0$ and require $f(\infty)=1$ or $-1,$ $h(\infty)=1$
or $-1,$ and $a(\infty)=1$ to obtain the correct asymptotic behavior.
We must also have $f=h=a=0$ at $r=0$ to have well-defined functions
at the origin. The equation $D_{j}c=0$ gives $f=h$ everywhere. Substituting
the ansatz into the expression for the energy, we obtain
\begin{eqnarray}
E_{BPS} & = & \mp\int dS_{j}B_{j}^{a}\left(\pm\right)\left(\frac{x^{a}}{r}w\cos\alpha+\frac{x^{a}}{r}v\sin\alpha\right)
\end{eqnarray}
where the $+$ sign in parenthesis corresponds to the case $f(\infty)=1$
and $-$ corresponds to $f(\infty)=-1.$ We identify $ $a magnetic
flux
\begin{equation}
\Phi=(\pm)\int dS_{j}B_{j}^{a}\frac{x^{a}}{r}=(\mp)\frac{4\pi}{g}
\end{equation}
and so the energy of the BPS monopole can be written as
\begin{eqnarray}
E_{BPS} & = & \mp\Phi\left(w\cos\alpha+v\sin\alpha\right).
\end{eqnarray}
Minimizing the energy as a function of $\alpha$, we obtain
\begin{equation}
E_{BPS}=\mp\Phi\sqrt{w^{2}+v^{2}}.
\end{equation}
By definition, $\Phi$ is negative for monopoles and positive for
antimonopoles. Thus the upper sign corresponds to monopole with $ $$f(\infty)=1$
and the lower sign to antimonopoles with $f(\infty)=-1$.

\begin{comment}
The first equation, $\left(D_{i}c\right)^{a}=0$, requires that $f=h$
everywhere. The remainder of the calculation is very similar to the
standard calculation for BPS monopoles, except that the vacuum expectation
value of the scalar field is $\sqrt{v^{2}+w^{2}}$ rather than $v$.
We define the dimensionless parameter 
\begin{equation}
\rho=g\sqrt{v^{2}+w^{2}}r
\end{equation}
 and find for the case of monopoles that the solutions are the well-known
BPS monopole solutions
\begin{eqnarray*}
a\left(\rho\right) & = & 1-\frac{\rho}{\sinh\rho}\\
f\left(\rho\right) & = & \coth\rho-\frac{1}{\rho}
\end{eqnarray*}
with a similar result for antimonopoles. Finally, we compute the magnetic
flux to be 
\begin{eqnarray*}
\Phi & = & \mp\frac{4\pi}{g}
\end{eqnarray*}
consistent with the Dirac charge quantization condition for an electric
charge $-g/2$. 
\end{comment}

In addition to the BPS monopole, there is another, topologically distinct
monopole which occurs at finite temperature when $A_{4}$ is treated
as a Higgs field. Starting from a static monopole solution where $\left|A_{4}\right|=w$
at spatial infinity, we apply a special gauge transformation
\begin{equation}
U_{special}=\exp\left[-\frac{i\pi x_{4}}{L}\tau^{3}\right]
\end{equation}
where $\tau^i$ is the Pauli matrix. Although $U_{special}$ is not periodic in $x^{4}$, it transforms
the scalar field as
\begin{equation}
\phi\rightarrow\exp\left[-\frac{i\pi x_{4}}{L}\tau^{3}\right]\phi\exp\left[+\frac{i\pi x_{4}}{L}\tau^{3}\right]
\end{equation}
 so that $\phi$ remains periodic: $\phi\left(\vec{x},x_{4}=0\right)=\phi\left(\vec{x},x_{4}=L\right)$.
However, $A_{\mu}$ transforms in such a way that the value of $A_{4}$
at spatial infinity is shifted: $w\rightarrow w-2\pi/gL$. If we instead
start from a static monopole solution such that $A_{4}=2\pi/gL-w$
at spatial infinity, then the action of $U_{special}$ gives a monopole
solution with $A_{4}=-w$ at spatial infinity. A final
constant gauge transformation $U_{const}=\exp\left[i\pi\tau^{2}/2\right]$
yields a new monopole solution with $A_{4}=w$ at spatial
infinity. The distinction between the BPS solution, which is independent
of $x_{4}$, and the KK solution is made clear by consideration of
the topological charge. The action of $U_{special}$ followed by $U_{const}$
increases the topological charge by $1$ and changes the sign of the
monopole charge. Thus the KK solution is topologically distinct from
the BPS solution because it carries instanton number $1$. This is
consistent with the KvBLL decomposition of instantons in the pure
gauge theory with nontrivial Polyakov loop behavior, where $SU(2)$
instantons can be decomposed into a BPS monopole and a KK monopole.
Our picture of the confined and mixed confined phases is one where
instantons and anti-instantons have {}``melted'' into their constituent
monopoles and antimonopoles, which effectively form a three-dimensional
gas of magnetic monopoles. In the BPS limit, both the magnetic and
scalar interactions are long-ranged; this behavior appears prominently,
for example, in the construction of $N$-monopole solutions in the
BPS limit. 

We thus find that the BPS solution has energy
\begin{equation}
E_{BPS}=\frac{4\pi}{g}\sqrt{w^{2}+v^{2}}
\end{equation}
 corresponding to an action
\begin{equation}
S_{BPS}=\frac{4\pi}{g}L\sqrt{w^{2}+v^{2}}=\frac{4\pi}{g^{2}}\sqrt{4\theta^{2}+g^{2}L^{2}v^{2}}.
\end{equation}
 For the KK solution, we have instead
\begin{equation}
S_{KK}=\frac{4\pi}{g^{2}}\sqrt{\left(2\pi-gLw\right)^{2}+g^{2}L^{2}v^{2}}=\frac{4\pi}{g^{2}}\sqrt{\left(2\pi-2\theta\right)^{2}+g^{2}L^{2}v^{2}}.
\end{equation}
Note that the action of a BPS monopole $S_{BPS}$ can be written in
the form $ML$, with $M$ independent of $L$. With $L$ regarded
as the inverse temperature $\beta$, this might suggest an interpretation
as a finite-energy solution of the Minkowski-space field equations.
However, the explicit presence of $\theta$ has no obvious Minkowski-space
interpretation. Furthermore, $S_{KK}$ cannot be written in the form
of a mass times $L$ in any case. This indicates that these monopole
solutions of the Euclidean field equations have no obvious continuation
to Minkowski, a point we shall reconsider in Sec.~VI.

\begin{comment}
NOTE: In the gauge where $A_{4}^{a}$ points in the $3$ direction
in group space at spatial infinity, we have for the eigenvalues of
$P$:
\begin{equation}
\exp\left[\pm igLw/2\right]
\end{equation}
 so $w=2\theta/gL$.
\end{comment}
{} 

Although we used the BPS construction to exhibit the existence and
some properties of the monopole solutions of our system, we must move
away from the BPS limit to ensure that magnetic interaction dominate
at large distances, \emph{i.e.}, that the three-dimensional scalar
interactions associated with $A_{4}$ and $\phi$ are not long-ranged.
This behavior is natural in the confined and mixed confined phases,
where the characteristic scale of the Debye (electric) screening mass
associated with $A_{4}$ is large, on the order of $g/L$. It is well
known that the BPS bound for the monopole mass holds as an equality
only when the scalar potential is taken to zero. As mentioned above,
in the case under consideration the scalar coupling $\lambda$ must
be very small for perturbation theory to be valid, but the potential
for $A_{4}$ is not small. However, the two combined potential can
be written together as a quartic potential in terms of the rotated
fields $b$ and $c$ with some quartic coupling $\lambda'$ for $b$.
Numerical studies \cite{Kirkman:1981ck} have shown that the monopole
action is given in general for $SU(2)$ as
\begin{equation}
LE_{BPS}C\left(\epsilon\right)
\end{equation}
where $\epsilon=\sqrt{\lambda'}/g$. The function $C\left(\epsilon\right)$
varies monotonically from $C\left(0\right)=1$ in the BPS limit to
$C\left(\infty\right)=1.787$ with the limiting behaviors 
\begin{equation}
C=1+\frac{\epsilon}{2}
\end{equation}
 and
\begin{equation}
C=1.787-\frac{2.228}{\epsilon}+O(\epsilon^{-2})
\end{equation}
for small and large $\epsilon$, respectively. Thus corrections to
the BPS result for the monopole mass and action due to the potential
terms are less than a factor of two. We will henceforth use the exact
results for the actions in the BPS limit, neglecting corrections from
$U_{eff}$ for the sake of simplicity of notation. It is useful to
note that the $SU(2)$ construction of the mixed phase monopoles extends
to $SU(N)$ in the standard way, via the embedding of $SU(2)$ subgroups
in $SU(N)$.

\section{Topological effects in the four phases}

We can now discuss the topological content of each of the four phases
we have found. It is important to understand that in all four phases,
Wilson loops in planes orthogonal to the compact direction should
show area-law behavior. This is an old observation about the deconfined
phase \cite{DeTar:1985kx,DeGrand:1986uf} which is very clearly observed
in lattice simulations of $SU(2)$ and $SU(3)$ at temperatures above
the deconfinement transition \cite{Bali:1993tz,Karsch:1994af}. At
first sight, this seems to directly conflict with the association
of deconfinement with the loss of area-law behavior for Wilson loops.
However, the introduction of a compact direction, as in the case of
finite temperature, explicitly breaks space-time symmetry. In the
case of finite temperature, Wilson loops measuring electric flux have
perimeter behavior in the deconfined phase; Wilson loops measuring
magnetic flux still obey an area law. This asymmetry in behavior can
be understood on the basis of center symmetry. The full center symmetry
of an $SU(N)$ gauge theory on a $d$-dimensional hypertorus $T^{d}$
is $Z(N)^{d}$. While the $Z(N)$ symmetry may break spontaneously
in the short compact direction, the other $Z(N)$ symmetries are unbroken,
and thus the associated Wilson loops obey an area law. Given the known
role of monopoles in the confined phase of $R^{3}\times S^{1}$ \cite{Unsal:2008ch},
it is in some sense unsurprising that monopoles might play a role
in the area law for Wilson loops in all four phases. 

In order to understand the effects of monopoles play in the four
phases we have identified, we must analyze their interactions. We
begin with a discussion of quantum fluctuations around the monopole
solutions. The contribution to the partition function of a single BPS monopole
at finite temperature was considered by Zarembo \cite{Zarembo:1995am},
and is given formally by
\begin{eqnarray}
Z{}_{a} & =\int d\mu^{a} & \exp\left[-S_{a}\right]\exp\left[-S_{d}\right]{\rm det'}\left[-\bar{D}_{\mu}^{2}+M_{g}^{2}\right]_{a}^{-1}{\rm det'}\left[-\bar{D}_{\mu}^{2}+M_{s}^{2}\right]_{a}^{-1/2}
\end{eqnarray}
where $a$ denotes the type of monopoles, $a=\left\{ BPS,KK,\overline{BPS},\overline{KK}\right\} $,
and the determinants are written with a prime to indicate that
zero modes are omitted.
The measure factor
$d\mu^{a}$ associated with the collective coordinates
(moduli) of the monopole solution, including the Jacobians from the
zero modes is given by \cite{Davies:2000nw} 
\begin{equation}
\int d\mu^{a}=\mu^{4}\int\frac{d^{3}x}{\left(2\pi\right)^{3/2}}J_{x}\int_{0}^{2\pi}\frac{d\phi}{\left(2\pi\right)^{1/2}}J_{\phi}
\end{equation}
where $x$ is the position and $\phi$ the $U(1)$ phase of the monopole
and $\mu$ is a Pauli-Villars regulator. The corresponding Jacobians
are
\begin{equation}
J_{x}=S_{a}^{3/2},\,\, J_{\phi}=NLS_{a}^{1/2}.
\end{equation}
Each of the four zero modes contributes a factor of $\mu$. We are
interested in the behavior of the model in the case where the eigenvalues
of $M_{s}^{2}$ and $M_{g}^{2}$ are much smaller than either $\mu^{2}$
or $L^{-2}$. For the functional determinants, this limiting case
is similar to the BPS limit, and the $\mu$ dependence of the
functional determinant is given by  \cite{Zarembo:1995am}
\begin{equation}
{\rm det'}\left[-\bar{D}_{\mu}^{2}+M_{s}^{2}\right]_{a}\approx{\rm det'}\left[-\bar{D}_{\mu}^{2}\right]_{a}\sim\left(NL\mu\right)^{1/3}
\end{equation}
for the scalar determinant and similarly for the gauge field determinant.
Collecting all the terms, each monopole carries a factor
\begin{eqnarray}
Z{}_{a} & = & c\mu^{7/2}\left(NL\right)^{1/2}S_{a}^{2}\exp\left[-S_{a}+\mathcal{O}\left(1\right)\right]\int d^{3}x \nonumber \\
 & = & \xi_{a}\exp\left[-S_{a}\right]\int d^{3}x
\end{eqnarray}
in its contribution to $Z$. The factor $\xi_{a}$ is $ $$c\mu^{7/2}\left(NL\right)^{1/2}S_{a}^{2}$
where $c$ is a numerical constant and the factor of $d^{3}x$ represents
the integration over the location of the monopole. From the construction
of the KK monopole, we see that we have $\xi_{KK}\left(\theta\right)=\xi_{BPS}\left(\pi-\theta\right)$.

\subsection{The Confined Phase}

The renormalization of the functional determinant arising from quantum
fluctuations around the monopole solution is particularly simple in
the confined phase, as first observed by Davies \emph{et al.} in the
corresponding supersymmetric model \cite{Davies:1999uw}. The dependence
on the Pauli-Villars regulator is removed, as usual, by coupling constant
renormalization. We begin by reviewing the previously studied cases
of a pure gauge theory with a deformation or with periodic adjoint
fermions. The relation at one-loop of the bare coupling and the regulator
mass $\mu$ to a renormalization-group-invariant scale $\Lambda$
is
\begin{equation}
\mbox{\ensuremath{\Lambda}}^{b_{0}}=\mu^{b_{0}}e^{-8\pi^{2}/g^{2}N}
\end{equation}
where $b_{0}$ is the first coefficient of the $\beta$ function divided
by $N$:
\begin{equation}
b_{0}=\frac{11}{3}-\frac{4}{3}\cdot\frac{n_{f}C(R_{f})}{N}-\frac{1}{6}\cdot\frac{n_{b}C(R_{b})}{N}
\end{equation}
where $n_{f}$ is the number of flavors of Dirac fermions in a representation
$R_{f}$, $n_{b}$ is the number of flavors of real scalars in a representation
$R_{b}$, and $C(R)$ is obtained from $Tr_{R}\left(T^{a}T^{b}\right)=C(R)\delta^{ab}$.
For the case of a pure gauge theory with a deformation, there are
four collective coordinates and this gives a factor of $\mu^{4}$.
The functional integral over gauge degrees of freedom gives rise to
a factor $\det'\left[-D^{2}\right]^{-1}\propto\mu^{-1/3}$ and the
action contributes a factor $\exp\left(-8\pi^{2}/g^{2}N\right)$
in the confined phase. Thus the contribution of a single monopole to
the partition function gives a factor
\begin{equation}
\mu^{4-\frac{1}{3}}e^{-8\pi^{2}/g^{2}N}=\mu^{11/3}e^{-8\pi^{2}/g^{2}N}=\Lambda^{11/3}.
\end{equation}
A detailed calculation confirms what we know on dimensional grounds:
the contribution $\xi_{a}e^{-8\pi^{2}/g^{2}N}\propto L^{-3}\left(\Lambda L\right)^{11/3}$.
Note that the eliminations of renormalization-dependent quantities
by renormalization-independent quantities depends crucially on the
coefficient of $1/g^{2}$ in the action. For the case of $n_{f}$
Dirac fermions in the adjoint representation, we have a factor
of $4 - 2 n_{f}$ from the zero modes: 
\begin{equation}
\mu^{(4-2n_{f})-\frac{1}{3}+2n_{f}\frac{1}{3}}e^{-8\pi^{2}/g^{2}N}=\mu^{11/3-4n_{f}/3}e^{-8\pi^{2}/g^{2}N}=\Lambda^{11/3-4n_{f}/3}
\end{equation}
for $n_{f}$ Dirac fermions which is again renormalization group invariant.

For a gauge theory with $n_{b}$ adjoint scalars plus a deformation,
we have similarly that
\begin{equation}
\mu^{11/3-n_{b}/6}e^{-8\pi^{2}/g^{2}N}=\Lambda^{11/3-n_{b}/6}.
\end{equation}
This implies that for $n_{b}=1$ the complete functional determinant
prefactor depends on $\Lambda$ and $L$ as $L^{-3}\left(\Lambda L\right)^{7/2}$.
As we have seen, the action of both the BPS and the KK monopole in
the gauge plus scalar model will exactly equal $8\pi^{2}/g^{2}N$ only in
the confined phase, so this result is special to that phase. 

%subsubsection{Interaction of monopoles and effective action}

The interaction of the monopoles is essentially the one described
by Polyakov in his original treatment of the Georgi-Glashow model
in three dimensions \cite{Polyakov:1976fu}, slightly generalized
to include both the BPS and KK monopoles. 
Let us consider, say, a BPS-type monopole and KK-type monopole located
at $\vec{x}_{1}$ and $\vec{x}_{2}$ in the noncompact directions,
with static worldlines in the compact direction. The interaction energy
due to magnetic charge of such a pair is
\begin{equation}
E_{BPS-KK}=-\left(\frac{4\pi}{g}\right)^{2}\frac{1}{4\pi\left|\vec{x}_{1}-\vec{x}_{2}\right|}
\end{equation}
and the associated action is approximately $S_{BPS}+S_{KK}+LE_{BPS-KK}$.
As discussed above, this will be larger than the value obtained from
the Bogomolny bound, but of the same order of magnitude. There is
an elegant way to capture the dynamics of the monopole plasma, using
an Abelian scalar field $\sigma$ dual to the magnetic field. Assuming
that the Abelian magnetic gauge field is three-dimensional for small
$L$, we may write
\begin{equation}
L\int d^{3}x\,\frac{1}{2}B_{k}^{2}=\int d^{3}x\frac{g^{2}}{32\pi^{2}L}\left(\partial_{k}\sigma\right)^{2}
\end{equation}
where the normalization of $\sigma$ is chosen to simplify the form
of the interaction terms. The three-dimensional effective action is
given by

\begin{equation}
L_{eff}=\frac{g^{2}}{32\pi^{2}L}\left(\partial_{j}\sigma\right)^{2}-\sum_{a}\xi_{a}e^{-S_{a}+iq_{a}\sigma}
\end{equation}
where the sum is over the set $\left\{ BPS,KK,\overline{BPS},\overline{KK}\right\} $.
Each species of monopole has its own magnetic charge sign $q_{a}=\pm$
as well as its own action $S_{a}$. The coefficients $\xi_{a}$ represent
the functional determinant associated with each kind of monopole,
but the combination $\xi_{a}\exp\left(-S_{a}\right)$ may be usefully
regarded as a monopole activity in terms of the statistical mechanics
of a gas of magnetic charges. The generating functional
\begin{equation}
Z_{\sigma}=\int\left[d\sigma\right]\exp\left[-\int d^{3}x\, L_{eff}\right]
\end{equation}
is precisely equivalent to the generating function of the monopole
gas. This equivalence is a generalization of the equivalence of a
sine-Gordon model to a Coulomb gas, and may be proved by expanding
$Z_{\sigma}$ in a power series in the $\xi_{a}$'s, and doing the
functional integral over $\sigma$ for each term of the expansion.

%\subsubsection{string tension}

It is well known that the magnetic monopole plasma leads to confinement
in three dimensions. For our effective three-dimensional theory, any
Wilson loop in a hyperplane of fixed $x_{4}$, for example a Wilson
loop in the $x_{1}-x_{2}$ plane, will show an area law. The original
procedure of Polyakov \cite{Polyakov:1976fu} may be used to calculate
the string tension, where the presence of a large planar Wilson loop
causes the dual field $\sigma$ to have a discontinuity on the surface
associated with the loop and a half-kink profile on both
sides.
However, an alternative procedure is simpler where the discontinuity
in the gauge field strength induced by the Wilson loop is moved to infinity
so that the string tension is obtained from the kink solution connecting
the two vacua of the dual field $\sigma$ \cite{Unsal:2008ch}.

In the confined phase, the action and functional determinant factors
for all four types of monopoles are the same, so we denote them by
$S_{M}$ and $\xi_{M}$. The potential term in the mixed and confined
phases then reduces to
\begin{equation}
-\sum_{a}\xi_{a}e^{-S_{a}+iq_{a}\sigma}\rightarrow4\xi_{M}e^{-S_{M}}\left[1-\cos\left(\sigma\right)\right]
\end{equation}
which has minima at $\sigma=0$ and $\sigma=2\pi$; we have added
a constant for convenience such that the potential is positive everywhere
and zero at the minima. A one-dimensional soliton solution $\sigma_{s}\left(z\right)$
connects the two vacua, and the string tension $\sigma_{3d}$ for
Wilson loops in the three noncompact directions is given by
\begin{equation}
\sigma_{3d}=\int_{-\infty}^{+\infty}dz\, L_{eff}\left(\sigma_{z}(z)\right)
\end{equation}
 which can be calculated via yet another Bogomolny inequality to be
\begin{equation}
\sigma_{3d}=\frac{4g}{\pi}\sqrt{\frac{\xi_{M}}{L}e^{-S_{M}}}.
\end{equation}
It is notable that in the confined phase $\sigma_{3d}$ can be written
in a form independent of the renormalization group scale.

\subsection{Generalization to Other Phases}

\begin{figure}
\includegraphics[width=5.5in]{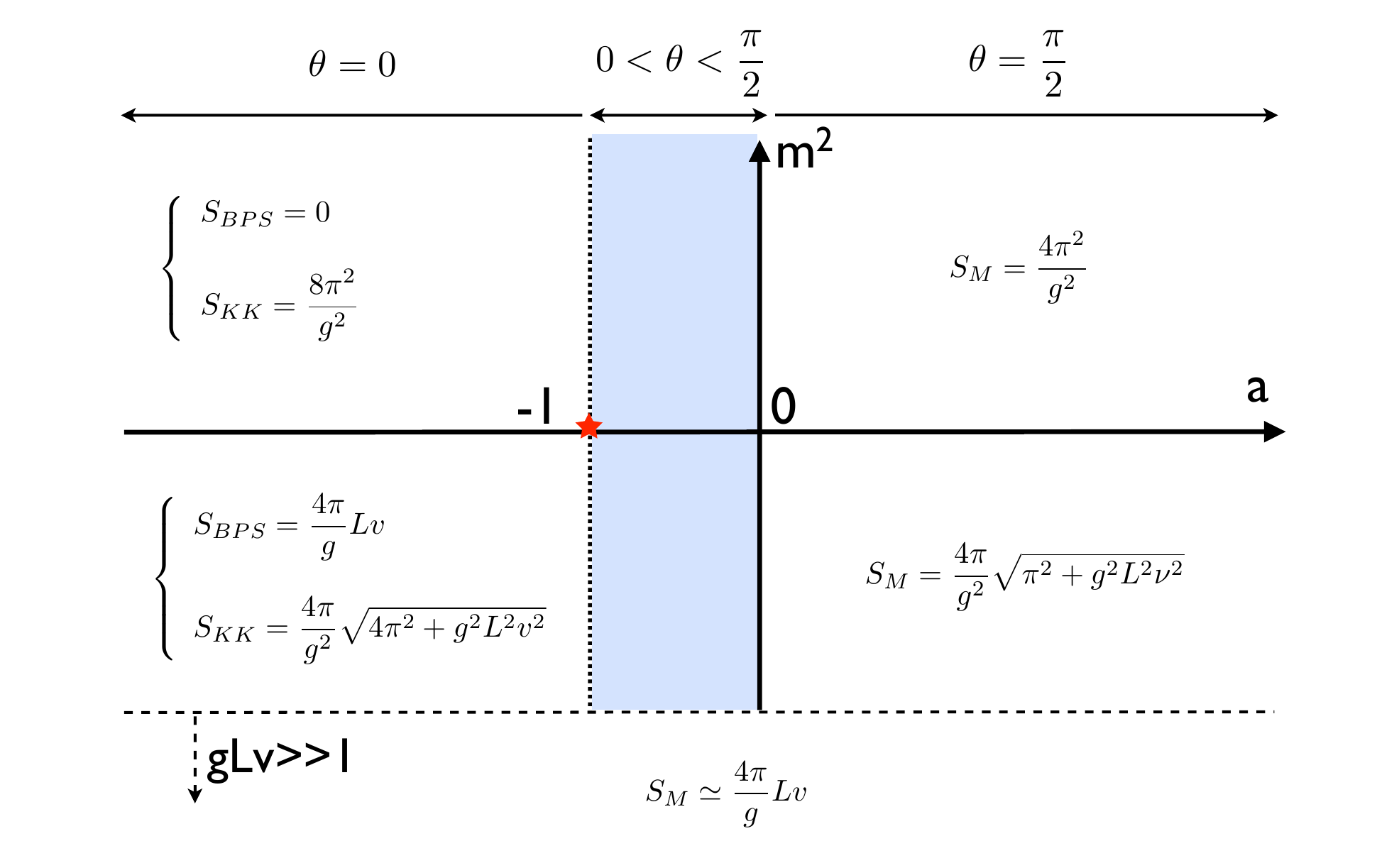}\caption{\label{fig:PhaseDiagramB}The phase diagram with the regions where
various limiting cases for $S_{BPS}$ and $S_{KK}$ hold; the shaded
region is a crossover region where $2>Tr_{F}P>0$. 
Crossover effects are negligible for $gLv\gg 1$.
The star marks
the point on the $m^{2}=0$ line where the deformation term is identically
zero.}
\end{figure}

The naive generalization of the above results for the confined phase
to the other three phases is straightforward. Writing explicitly the
$\theta$ dependence, we have in general that $S_{BPS}\left(\theta\right)$
is not the same as $S_{KK}\left(\theta\right)$, and for arbitrary
$\theta$, $\xi_{BPS}\left(\theta\right)\ne\xi_{KK}\left(\theta\right)$.
However, it is generally true that $\xi_{BPS}\left(\theta\right)=\xi_{\overline{BPS}}\left(\theta\right)$
and $\xi_{KK}\left(\theta\right)=\xi_{\overline{KK}}\left(\theta\right)$;
furthermore, the explicit construction of the KK monopole from the
BPS monopole shows that $\xi_{BPS}\left(\theta\right)=\xi_{KK}\left(\pi-\theta\right)$.
The limiting cases of $S_{BPS}$ and $S_{KK}$ for $\theta=0$ and
$\pi/2$ and for $v=0$ and large $v$ are shown in Fig.~\ref{fig:PhaseDiagramB}.

The construction of the sine-Gordon dual Lagrangian proceeds in a
familiar way. Essentially, we must make the replacement
\begin{equation}
\xi_{BPS}\left(\pi/2\right)e^{-S_{BPS}\left(\pi/2\right)}\rightarrow\frac{1}{2}\left(\xi_{BPS}\left(\theta\right)e^{-S_{BPS}\left(\theta\right)}+\xi_{KK}\left(\theta\right)e^{-S_{KK}\left(\theta\right)}\right).
\end{equation}
Repeating the calculation of the string tension leads to
\begin{equation}
\sigma_{3d}=\frac{4g}{\pi}\sqrt{\frac{1}{2L}\left(\xi_{BPS}\left(\theta\right)e^{-S_{BPS}\left(\theta\right)}+\xi_{KK}\left(\theta\right)e^{-S_{KK}\left(\theta\right)}\right)}.
\end{equation}
However, there are two issues raised by this generalization.
The first is the validity of the dilute monopole gas approximation.
The assumption that the monopoles can be treated as well-separated
objects will hold when $\xi_{a}\exp\left(-S_{a}\right)\ll1$ for all
monopole species. We will examine this point in detail below for all
three remaining phases. 

The second issue is technical: the renormalization group-dependence
of the final result for $\sigma_{3d}$. As we have seen, in the confined
phase $\sigma_{3d}$ can be written in terms of $L$ and $\Lambda$,
with no dependence on the regulator $\mu$.
When $\theta\ne\pi/2$, the explicit cancellation of the $\mu$ dependence
between $\xi_{a}$ and $\exp\left(-S_{a}\right)$ does not occur:
the $\mu$ dependence of $\xi_{a}\left(\theta\right)$ does not depend
on $\theta$, but the coefficient of $1/g^{2}\left(\mu\right)$ in
$S_{a}$ is $\theta$-dependent. 
This issue is not new, and
not specific to Higgs models; it
was discussed in the supersymmetric case in \cite{Davies:1999uw}
in the context of the effective potential for $\sigma$.
However, the effective Lagrangian $L_{eff}$ represents only the
long-distance behavior of the model; in fact, the cosine
interaction is not even renormalizable in three dimensions.
The underlying gauge theory is of course renormalizable,
and the ultraviolet renormalization of instanton effects
is well-understood. In the case of pure gauge theory, the
renormalizability of monopole gas effects has been
confirmed by detailed analysis of the relevant functional
determinants \cite{Diakonov:2004jn,Diakonov:2005qa}.
On the other hand,
the effective Lagrangian represents only
the long-ranged interaction mediated by $\sigma$.
This interaction falls off very slowly with distance,
because it is induced by nonperturbative effects.
Interactions mediated by particles with masses obtained
from perturbation theory must be integrated out
to obtain $L_{eff}$ \cite{Unsal:2008ch}.
This induces a dependence of the parameters of $L_{eff}$
on some intermediate momentum scale on
the order of the lightest perturbative mass.
In the case at hand, this will be either the mass associated
with $A_4^3$ or $\phi^3$, which are obtained 
by minimizing the effective potential with respect
to $\theta$ and $v$. Thus $L_{eff}$ is only
valid up to the lightest perturbative scale,
and its finite parameters depend implicitly on that scale,
which in turn depend on $\theta$ and $v$.
Thus the monopole activities are not
simple functional determinants, but include
the effects of integrating the instanton gas
down to a scale where only the
$\sigma$ interaction remains.
For notational simplicity,
we will continue to denote
the monopole activities by
$\xi_a(\theta) \exp\left( -S_a(\theta)\right)$.
We now turn to consideration of the
deconfined, mixed confined and Higgs phase in turn.

%OLD v 1
%The second issue is the renormalization group-dependence
%of the final result for $\sigma_{3d}$. As we have seen, in the confined
%phase $\sigma_{3d}$ can be written in terms of $L$ and $\Lambda$,
%with no dependence on a renormalization scale $\mu$.
%When $\theta\ne\pi/2$, the cancellation of the $\mu$ dependence
%between $\xi_{a}$ and $\exp\left(-S_{a}\right)$ does not occur:
%the $\mu$ dependence of $\xi_{a}\left(\theta\right)$ does not depend
%on $\theta$, but the coefficient of $1/g^{2}\left(\mu\right)$ in
%$S_{a}$ is $\theta$-dependent. Thus in general we will have an expression
%for a physical quantity, $\sigma_{3d}$, that depends on $\mu$. This
%is unpleasant, but not unprecedented. For example, the perturbative
%calculation of the pressure of the quark-gluon plasma shows a $\mu$
%dependence. It is perhaps unsurprising that a semiclassical calculation
%might show a similar dependence. We now turn to consideration of the
%deconfined, mixed confined and Higgs phase in turn.

In the deconfined phase, we have $v=0$, but $Z(2)_{C}$ is broken
so $\theta\ne\pi/2$ and $Tr_{F}P\ne0$. As we cross from the confined
to the deconfined phase, the second-order character of the deconfinement
transition means that $\theta$ will move continuously from its $Z(2)_{C}$
-symmetric value of $\pi/2$ towards $0$ as $a$ is decreased below
$0$. Throughout this phase, $v=0$ and thus we have for the BPS action
\begin{equation}
S_{BPS}=\frac{4\pi}{g^{2}}\cdot2\theta
\end{equation}
and for the KK solution, we have instead
\begin{equation}
S_{KK}=\frac{4\pi}{g^{2}}\left(2\pi-2\theta\right).
\end{equation}

There is a natural region in the deconfined phase where the monopole
dynamics is essentially identical to that in the confined phase. We
begin by expanding the monopole actions around $\theta=\pi/2$. The
BPS action in this limit becomes
\begin{equation}
S_{BPS}=\frac{8\pi}{g^{2}}\theta=\frac{4\pi^{2}}{g^{2}}+\frac{8\pi}{g^{2}}\delta
\end{equation}
where we have made the substitution $\theta=\pi/2+\delta$. The KK
action becomes in the same limit
\begin{equation}
S_{KK}=\frac{4\pi}{g^{2}}\left(2\pi-2\theta\right)=\frac{4\pi^{2}}{g^{2}}-\frac{8\pi}{g^{2}}\delta.
\end{equation}
In order to obtain monopole physics similar to that of the confined
phase we must require 
\begin{equation}
S_{BPS}=S_{KK}=\frac{4\pi^{2}}{g^{2}}+\mathcal{O}\left(1\right)
\end{equation}
which in turn implies that $\delta$ is no larger than $\mathcal{O}\left(g^{2}\right)$.
From the effective action we constructed in Sec.~III, we have
in the deconfined phase
\begin{equation}
U_{eff}=\frac{2}{\pi^{2}L^{4}}\delta^{4}+\frac{a}{L^{4}}\delta^{2}
\end{equation}
we see that $\delta$ will be nonzero only if $a$ is negative. In
that case, we must have 
\begin{equation}
\left|a\right|\propto\delta^{2}\lesssim g^{4}.
\end{equation}
Thus the approximation that, $S_{BPS}=S_{KK}=\frac{4\pi^{2}}{g^{2}}+\mathcal{O}\left(1\right)$,
is valid only in a very narrow region in the deconfined phase where $\theta=\pi/2-\mathcal{O}(g^{2})$
and $ $$\left|a\right|\lesssim g^{4}$. We also expect that the functional
determinants of the BPS and KK monopoles are approximately equal in
this region. Thus, in this region all of the monopole physics which
we worked out for the confined phase is valid: the monopole plasma
is equivalent to a sine-Gordon field theory, and the string tension
is obtained from the sine-Gordon kink solution.

In the region where $a<-1$, $\theta$ is zero, and we know that the
interpretation of a finite-temperature instanton in terms of monopole
constituents is probably lost. In pure gauge theories, the monopole
constituent picture of the instanton breaks down at the classical
level when $\theta\rightarrow0$. As shown in \cite{Kraan:1998kp,Kraan:1998pm},
in the pure $SU(2)$ gauge theory the instanton action density is
well-localized into two separate lumps when $\theta=\pi/$2, but only
one lump persists when $\theta\rightarrow0$. This is reflected in
the behavior of the formula for $S_{BPS}$ as $\theta$ approaches
zero. Nevertheless, the total action of a BPS-KK pair stays exactly
at $S_{instanton}=S_{BPS}+S_{KK}=8\pi^{2}/g^{2}$ for all values of
$\theta$. This suggests that the bulk of the confined phase, where
$a<-1$, might be best interpreted in terms of an instanton gas rather
than as a gas of monopoles. This region would then naturally extend
to the right of the line segment $a=-1$ by a factor of $\mathcal{O}\left(g^2\right)$.
However, it should be noted that the work
of Rossi \cite{Rossi:1978qe} showed that for pure gauge theories
with $\theta=0$, an infinite line of four-dimensional instantons
with spacing $L$ and scale parameter $2\pi/L$ is exactly equivalent
to a monopole solution of the field equations. This solution was later
realized to be equivalent to the $\theta=0$ limit of the KK monopole.
We will return to the relation of the Euclidean and Minkowski solutions
below when we discuss a certain duality present in the system. The
region where $0<\theta<\mbox{\ensuremath{\pi}/2}$, corresponding
to $-1<a<0$, appears to be a crossover region where the interpretation
of the topological content is not yet clear, as the system moves smoothly
from a dilute monopole gas near $a=0$ to a phase where $S_{BPS}=0$
for $a\le-1$.

In the mixed phase, $\theta=\pi/2$ and $v$ is nonzero. We have
\begin{equation}
S_{BPS}=S_{KK}=\frac{4\pi}{g^{2}}\sqrt{\pi^{2}+g^{2}L^{2}v^{2}}
\end{equation}
 as in the confined phase. The functional determinants $\xi_{BPS}\left(\pi/2\right)$
and $\xi_{KK}\left(\pi/2\right)$ are equal as well. Because $v \ne 0$,
the handling of ultraviolet divergences is not as simple as in the
confined phase, but can be carried
out in principle \cite{Kiselev:1988gf,Kiselev:1990fh}. 
The analysis of the string tension performed for
the confined phase carries over, and $\sigma_{3d}$ is given by
\begin{equation}
\sigma_{3d}=\frac{4g}{\pi}\sqrt{\frac{\xi_{M}}{L}e^{-S_{M}}}
\end{equation}
where as before $S_{M}$ and $\xi_{M}$ are the common monopoles in
this phase. There is a natural region next to the confined phase where
$gLv<\pi$. In that region, we again have $S_{M}=4\pi^{2}/g^{2}+\mathcal{O}\left(1\right)$
and the renormalization group arguments used in the confined phase
work here as well. Although not natural in the case $g(L)\ll1$, there
is a region far from the confined region where $gLv\gg1$, where $S_{M}\approx4\pi Lv/g$.
This is precisely the action of a Minkowski-space monopole of mass
$4\pi v/g$ with a worldline of length $L$; we return to this point
in the discussion of duality below.

\begin{figure}
\includegraphics[width=5.5in]{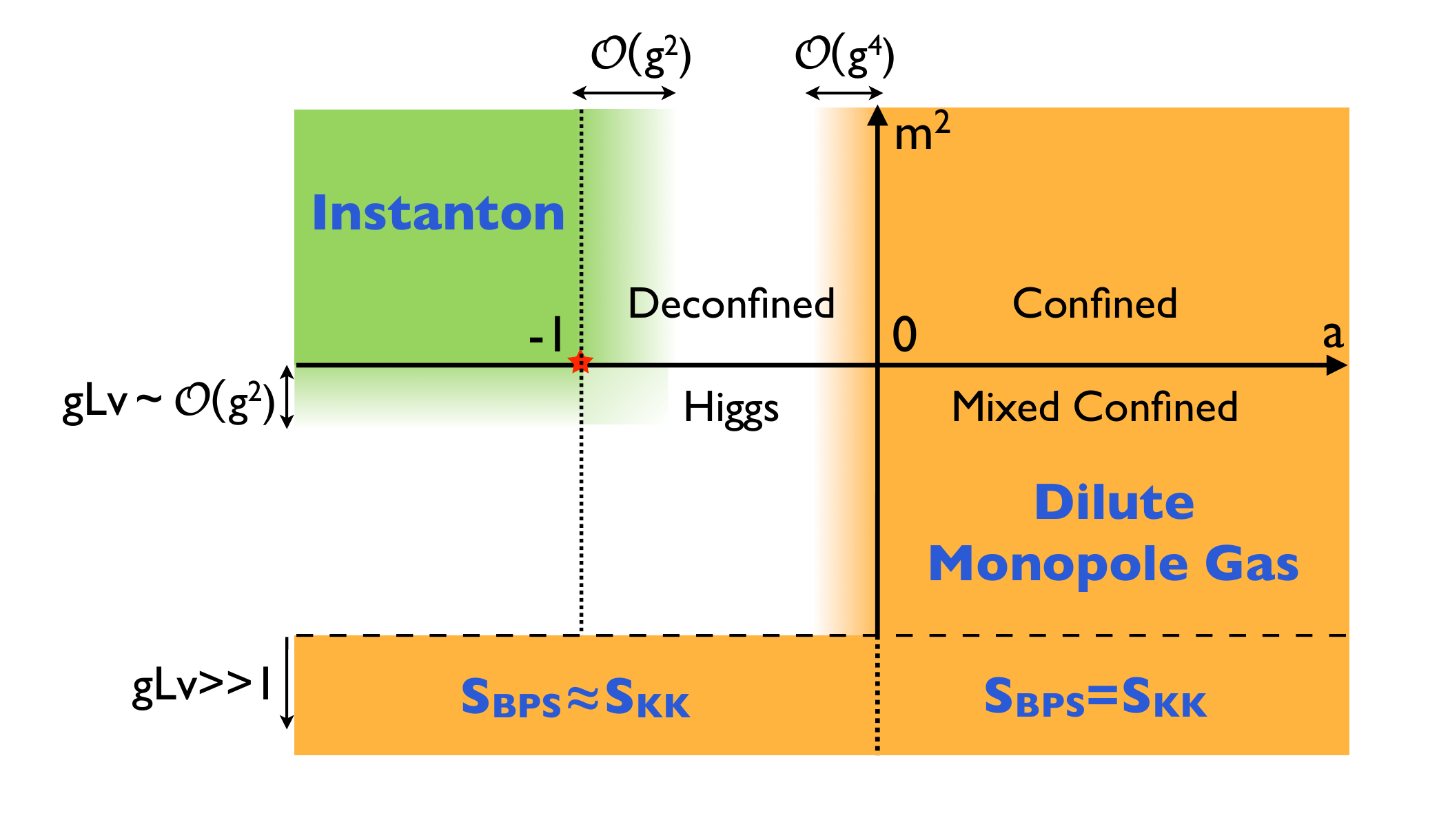}\caption{\label{fig:PhaseDiagramC}The phase diagram showing the region where the dilute monopole gas approximation is valid and
$S_{BPS}\approx S_{KK}$. The dilute gas region itself is somewhat
larger than the shaded region. The region labeled instanton is where
the dilute monopole picture does not hold. }
\end{figure}

In the Higgs phase, we have $\theta\ne\pi/2$ and $v \ne 0$
so both $Z(2)$ symmetries are broken. The action of a BPS monopole
solution is
\begin{equation}
S_{BPS}=\frac{4\pi}{g^{2}}\sqrt{4\theta^{2}+g^{2}L^{2}v^{2}}
\end{equation}
 but for the KK solution, we have instead
\begin{equation}
S_{KK}=\frac{4\pi}{g^{2}}\sqrt{\left(2\pi-2\theta\right)^{2}+g^{2}L^{2}v^{2}}.
\end{equation}
There are several regions of interest with the Higgs phase. Near the
critical line where $\theta$ is close to $\pi/2$, the behavior is
similar to that of the mixed confined phase; the argument is exactly
the same as for the deconfined phase when $\theta\approx\pi/2$ in
relation to the confined phase. We also expect behavior similar to
that of the mixed confined phase when $gLv\gg1$. 
The region $gLv\gg1$ may be treated in a manner very similar
to Polyakov's original treatment of the three-dimensional Georgi-Glashow
model, except that there is an additional factor of $2$ in the monopole
fugacity, and the three-dimensional instanton action $S_{3d}$ is
replaced by $4\pi Lv/g$.
In both these regions,
we have the approximate equality $S_{BPS}\approx S_{KK}$, and we
expect the dilute monopole gas picture is valid. There is also a region
where $gLv\le\mathcal{O}\left(g^{2}\right)$ and $\theta=0$ (for
$a<-1$) which has the behavior of the $\theta=0$ region of the deconfined
phase. 

In Fig.~\ref{fig:PhaseDiagramC}, we show a final version of the
phase diagram. The figure shows the large region where the dilute
monopole gas description should be valid, and either $S_{BPS}=S_{KK}$
or $S_{BPS}\simeq S_{KK}$. Note that this region includes all of
the confined and mixed confined regions, a large part of the Higgs
phase, and a small part of the deconfined phase. The region where
the dilute gas approximation is valid is somewhat larger. However,
we have also indicated the region where the dilute gas approximation
breaks down, because $S_{BPS}\approx0$ and $S_{KK}\approx8\pi^{2}/g^{2}$.
For obvious reasons, we have labeled this region as an instanton region,
although the correct treatment of topological excitations in this
region is no clearer in the Higgs system than in the pure gauge case.

\subsection{Duality}

As we have seen, the regions where various approximations hold are
not necessarily coincident with the phase boundaries. Essentially,
the mixed confined phase mediates between the confined and Higgs phases,
producing a broad band where the confining behavior of Wilson loops
can be ascribed to a dilute monopole gas. Across each phase boundary
(except possibly for the Higgs-deconfined boundary), the semiclassical
expression for the string tension measured by Wilson loops varies
smoothly. This would not be expected if the phase transitions were
first-order, and singular corrections to the semiclassical
picture are possible for second-order transitions due to coupling
between the order parameters and the dual field $\sigma$. This sort
of coupling of different order parameters is familiar in the PNJL
model \cite{Nishimura:2009me}. More important than the smooth behavior
of the string tension, however, is the continuity of the monopole
confinement mechanism across the confined, mixed and Higgs phases.

We can understand the role of topological excitations from
a different point of view
by invoking duality in a form similar to that used by Poppitz and
Unsal in their analysis of the Seiberg-Witten model \cite{Poppitz:2011wy};
their work also serves as an introduction to duality in this context.
The general issue in their work and here is the relation between
topologically-stable solutions of the classical field equations
in Euclidean space and Minkowski space. 
These are respectively solutions with finite action (instantons) and
finite energy (monopoles).
Higgs models
with adjoint scalars have both, and two different approaches
for computing the partition function on $R^3 \times S^1$ suggest
themselves. We have extensively discussed the use of instantons,
but another approach would be to consider the statistical mechanics
of Minkowski-space solutions with finite energy, which are monopoles
or more generally Julia-Zee dyons \cite{Julia:1975ff}. Such dyons will make
contributions to the overall partition function proportional to
$ \exp\left( -LM \right) P$, where $M$ is the monopole mass
and $P$ is a Polyakov loop factor.
As we will see below, there is evidence that summing over
finite-action instanton contributions to the partition function is equivalent
to summing over finite-energy dyon contributions,
extending the ideas in \cite{Poppitz:2011wy} 
to the case of nonsupersymmetric Higgs models on $R^3 \times S^1$.

Our approach is somewhat different from that of Poppitz and Unsal,
in that we relate a finite rather than infinite sum
over Euclidean monopoles to an infinite sum of Minkowski-space dyons.
We begin with an easy variant of the Poisson summation formula associated
with $Z(N)_{C}$. Let $f(\theta)$ be a function defined on the interval
$-\pi<\theta<\pi$. We define the Fourier series in the usual way:
\begin{eqnarray}
f\left(\theta\right) & = & \sum_{n\in Z}\tilde{f}\left(n\right)e^{in\theta}\\
\tilde{f}\left(n\right) & = & \int_{-\pi}^{\pi}\frac{d\theta}{2\pi}f\left(\theta\right)e{}^{-in\theta}.
\end{eqnarray}
Then we have that
\begin{eqnarray}
\sum_{k=0}^{N-1}f\left(\theta-\frac{2\pi k}{N}\right) & = & \sum_{n\in Z}\tilde{f}\left(n\right)\sum_{k=0}^{N-1}e^{in\left(\theta-\frac{2\pi k}{N}\right)} \nonumber \\
 & = & \sum_{n\in Z}\tilde{f}\left(n\right)e^{in\theta}N\delta\left(n\equiv0\left(N\right)\right)\nonumber \\
 & = & \sum_{n\in Z}N\tilde{f}\left(nN\right)e^{inN\theta}
\end{eqnarray}
so that for $N=2$ only the even coefficients $\tilde{f}(2n)$ contribute.
Let us apply this identity to the combination
\begin{equation}
\xi_{BPS}\left(\theta\right)e^{-S_{BPS}\left(\theta\right)}+\xi_{KK}\left(\theta\right)e^{-S_{KK}\left(\theta\right)}=\xi_{BPS}\left(\theta\right)e^{-S_{BPS}\left(\theta\right)}+\xi_{BPS}\left(\pi-\theta\right)e^{-S_{BPS}\left(\pi-\theta\right)}
\end{equation}
 which occurs in the dual Lagrangian and in the formula for $\sigma_{3d}$
so we have
\begin{equation}
f\left(\theta\right)=\xi_{BPS}\left(\theta\right)e^{-S_{BPS}\left(\theta\right)}.
\end{equation}
 For small $g^{2}$, $S_{BPS}\left(\theta\right)$ is strongly peaked
at $\theta=0$, so we can make the approximation
\begin{eqnarray}
\tilde{f}\left(n\right) & \simeq & \int_{0}^{\infty}\frac{d\theta}{\pi}\xi_{BPS}\left(0\right)e^{-S_{BPS}\left(\theta\right)}e^{in\theta}.
\end{eqnarray}
 Although this integral, with the limits taken to infinity, can be
evaluated in a saddle point approximation, it can also be evaluated
exactly \cite{Poppitz:2011wy}, giving
\begin{equation}
\tilde{f}\left(2n\right)\simeq\xi_{BPS}\left(0\right)\frac{gLv}{2\pi}\cdot\frac{\frac{4\pi}{g^{2}}}{\sqrt{\left(\frac{4\pi}{g^{2}}\right)^{2}+n^{2}}}K_{1}\left[gLv\sqrt{\left(\frac{4\pi}{g^{2}}\right)^{2}+n^{2}}\right].
\end{equation}

The Higgs phase represents the most general domain of applicability
of the duality transformation, because in the Higgs phase $v\ne0$
and $0\le\theta<\pi/2$. It is natural to introduce $M(n)$ the mass
of a Minkowski-space Julia-Zee dyon \cite{Julia:1975ff}
of magnetic charge $4\pi/g$ and
electric charge $ng$
\begin{equation}
M\left(n\right)=gv\sqrt{\left(\frac{4\pi}{g^{2}}\right)^{2}+n^{2}}
\end{equation}
 so that we can write
\begin{equation}
\tilde{f}\left(2n\right)\simeq\xi_{BPS}\left(0\right)\frac{LM(0)}{2\pi}\cdot\frac{1}{\sqrt{\left(\frac{4\pi}{g^{2}}\right)^{2}+n^{2}}}K_{1}\left[LM(n)\right].
\end{equation}
 The asymptotic expansion of the Bessel function for large argument
gives a factor of $\exp\left[-LM(n)\right]$:
\begin{equation}
\tilde{f}\left(2n\right)\simeq\xi_{BPS}\left(0\right)\frac{LM(0)}{2\pi}\cdot\frac{1}{\sqrt{\left(\frac{4\pi}{g^{2}}\right)^{2}+n^{2}}}\sqrt{\frac{\pi}{2LM(n)}}\exp\left[-LM(n)\right].
\end{equation}
Thus each term in the sum carries a factor of $\exp\left[-LM(n)+i2n\theta\right]$.
This suggests an obvious interpretation of the finite sum over BPS
and KK monopoles, which are constituents of instantons, as being equivalent
to a gas of Julia-Zee dyons, each carrying a Polyakov loop
factor appropriate to its charge. This interpretation is valid throughout
most of the Higgs and mixed confined phases, except in the region
near $m^{2}=0$ where the mass of the lightest dyon $M(0)=4\pi v/g$,
which is a Minkowski-space monopole, becomes light. Within this framework, 
the only significant difference between the mixed confined
and Higgs phases is that in the mixed confined phase, $\theta$ is
restricted to $\pi/2$.

When we cross the phase boundary $m^{2}=0$, we move into a region
where $v=0$. As long as we stay away from the region where $\theta$
is zero or $\mathcal{O}\left(g^{2}\right)$, the approximate form
of the Fourier coefficients is valid, and we have 
\begin{equation}
\tilde{f}\left(2n\right)\simeq\frac{2g^{2}}{16\pi^{2}+g^{4}n^{2}}
\end{equation}
which tells us that
\begin{equation}
\exp\left(-\frac{8\pi}{g^{2}}\theta\right)+\exp\left(-\frac{8\pi}{g^{2}}\left(\pi-\theta\right)\right)\approx\sum_{n\in Z}\frac{4g^{2}}{16\pi^{2}+g^{4}n^{2}}e^{i2n\theta}.
\end{equation}
Although the right-hand side is a good approximation to the left-hand
side as $\theta$ varies, it is striking how different the two forms
are. However, an exact evaluation of $\tilde{f}\left(2n\right)$ in
the limit $v=0$ gives
\begin{eqnarray}
\tilde{f}\left(2n\right) & = &{1\over \pi} \int_{0}^{\pi}{d\theta}\exp\left[-\left(\frac{8\pi}{g^{2}}+i2n\right)\theta\right]=\frac{1}{\frac{8\pi}{g^{2}}+i2n}\left(1-e^{-8\pi^{2}/g^{2}}\right)
\end{eqnarray}
showing that the nonperturbative behavior has not totally disappeared.

\section{Conclusions}

We have shown that the phase structure of the deformed $SU(2)$ adjoint
Higgs model on $R^{3}\times S^{1}$ is rich, with four different phases
distinguished by the behavior of the three gauge-invariant order parameters
associated with the global symmetries of the model. We have used a
particular deformation which makes the phase diagram simple, but the
appearance of four distinct phases is general. Despite the $Z\left(2\right)_{C}\times Z\left(2\right)_{H}$
global symmetry, the phase transitions separating the different phases
may be of second order or of first order. In addition to the known
confined, deconfined and Higgs phases, we have found a fourth phase,
the mixed confined phase, which takes the place of what would be a
confining phase with a Higgs mechanism. In the mixed confined phase,
the behavior of $A_{4}$ and $\phi$ become entwined in such a way
that the global symmetry group $Z(2)_{C}\times$$Z(2)_{H}$ breaks
spontaneously to a $Z(2)$ symmetry which acts nontrivially on both
$A_{4}$ and $\phi$. This behavior, found using perturbation theory,
extends to the topological properties of the model, where the BPS
and KK monopole solutions are constructed using a linear combination
of $A_{4}$ and $\phi$. The area-law behavior of Wilson loops orthogonal
to the compact $S^{1}$ direction can be attributed to a dilute magnetic
monopole gas in at least part of all four phases. There are several
unresolved issues. The correct treatment of topology in the deconfined
phase when $\theta=0$, corresponding to the high-temperature limit
$T\gg\Lambda$ in the case of finite temperature, remains elusive.
A detailed calculation of the monopole activities in the effective
Lagrangian which determines $\sigma_{3d}$
would be useful in comparing with lattice results.
The correct interpretation of the duality between Euclidean-space
monopoles, which are constituents of monopoles, and Minkowski-space
dyons is compelling, but incomplete. There is also the question
of generalizing our $SU(2)$ results to $SU(N)$ adjoint Higgs models
on $R^{3}\times S^{1}$. 
For $SU(N)$ gauge theories on $R^3\times S^1$, the natural set
of order parameters is $Tr_F P^k$, and the $Z(N)$ center symmetry
can break to a subgroup $Z(p)$ \cite{Myers:2007vc,Myers:2009df}. 
With the addition of an adjoint
scalar, there is the additional set of order parameters of the form
$Tr_F P^k\phi$ available. This suggests a very rich
phase structure is possible.
Finally, we note that many of the predictions we have made
may be difficult to test, because lattice simulations
of the three-dimensional Higgs model are consistent
with Polyakov's semiclassical results only over ad
narrow region \cite{Wensley:1989ja}.
However, the overall phase structure we have
predicted in our four-dimensional model
should be relatively easy to test with lattice simulations.

\acknowledgments{The authors thank the U.S.~Department of Energy for support.}

\bibliography{Higgs_vs_Confinement_ms_2A}

\end{document}